\documentclass[fleqn,usenatbib]{mnras}

\usepackage{newtxtext}
\usepackage[varvw]{newtxmath}
\usepackage{gensymb}

\usepackage[T1]{fontenc}

\DeclareRobustCommand{\VAN}[3]{#2}
\let\VANthebibliography\thebibliography
\def\thebibliography{\DeclareRobustCommand{\VAN}[3]{##3}\VANthebibliography}

\usepackage{graphicx}	% Including figure files
\usepackage{amsmath}	% Advanced maths commands
\usepackage{bm}
\usepackage{pgfgantt}
\usepackage[cal = cm]{mathalpha}
\usepackage{tikz}
\usetikzlibrary{shapes,arrows,fit}
\usepackage{braket}
\usepackage{booktabs,tabularx,multirow}
\newcolumntype{C}{>{\centering\arraybackslash}X}
\newcolumntype{Y}{>{\raggedright\arraybackslash}X}
\newcolumntype{S}{>{\centering\arraybackslash}m{1.0cm}}
\newcolumntype{V}{>{\centering\arraybackslash}m{2.7cm}}
\usepackage{pifont}
\usepackage[normalem]{ulem}
\newcommand{\rev}[1]{#1}

\newcommand{\anoise}{\ensuremath{\alpha\textit{Noise}}}

\newcommand{\mwedge}{\ensuremath{\textit{Wedge}}}

\newcommand{\manalytic}{\ensuremath{\textit{Analytic}}}

\newcommand{\avarl}{\ensuremath{\alpha\textit{VarL}}}

\newcommand{\ubase}{\ensuremath{\textit{Unibaseline}}}

\title[Bayesian GPR models]{Bayesian model comparison and validation with Gaussian Process Regression for interferometric 21-cm signal recovery}

\author[Y. Liu et al.]{
Yuchen Liu,$^{1,2}$\thanks{E-mail: yl871@cam.ac.uk}
Eloy de Lera Acedo,$^{1,2}$
Peter H. Sims$^{1,2}$
\\
$^{1}$Cavendish Astrophysics, University of Cambridge, JJ Thomson Avenue, Cambridge CB3 0HE, UK\\
$^{2}$Kavli Institute for Cosmology,
University of Cambridge, Madingley Road, Cambridge CB3 0HA, UK\\
}

\date{Accepted XXX. Received YYY; in original form ZZZ}

\pubyear{\the\year{}}

\begin{document}
\label{firstpage}
\pagerange{\pageref{firstpage}--\pageref{lastpage}}
\maketitle

% list models from different metrics

\begin{abstract} 
The 21-cm signal from neutral hydrogen traces the formation and evolution of early cosmic structures during the Cosmic Dawn and the subsequent Epoch of Reionization. However, the intrinsic faintness of the signal, as opposed to astrophysical foregrounds, poses a formidable challenge for its detection. Motivated by the recent success of machine learning based Gaussian Process Regression (GPR) methods in LOFAR and NenuFAR observations, we perform a Bayesian comparison among five GPR models to account for simulated 4-hour tracking observations with the SKA-Low telescope. The simulations incorporate the beam response of the telescope and include realistic radio sources and thermal noise from 122 to 134 MHz. A Bayesian model evaluation framework is applied to five GPR models to discern the most effective modelling strategy and determine the optimal model parameters. The GPR model with wedge parametrization (\mwedge{}) and its extension (\anoise{}) with noise scaling achieve the highest Bayesian evidence of the observed data and the least biased 21-cm power spectrum recovery. The \mwedge{} and \anoise{} models also forecast the best local power-spectrum recovery, demonstrating fractional differences of $0.10\%$ and $-0.24\%$ respectively, compared to the injected 21-cm power at $k = 0.32\ \mathrm{h\ cMpc}^{-1}$. We additionally perform Bayesian null tests to validate the five models, finding that the two optimal models also pass with the remaining three models yielding spurious detections in data containing no 21-cm signal.
\end{abstract}

% Bayesian model selection in abstract
% list models in abstract
% adding more comparison numbers

\begin{keywords}
methods: data analysis -- methods: statistical -- techniques: interferometric -- dark ages, reionization, first stars -- cosmology: observations
\end{keywords}

%%%%%%%%%%%%%%%%%%%%%%%%%%%%%%%%%%%%%%%%%%%%%%%%%%

%%%%%%%%%%%%%%%%% BODY OF PAPER %%%%%%%%%%%%%%%%%%

\section{Introduction}

The 21-cm transition of atomic hydrogen is a crucial observational tool in radio astronomy to trace the astrophysical processes during the Cosmic Dawn and the Epoch of Reionization (EoR). Given the abundance of hydrogen in the Universe, the signal provides insights into the formation of the early-formed structures, such as Pop III stars, along with their evolutionary path across $6 < z < 30$. Furthermore, it allows us to explore how their radiation influences the surrounding intergalactic medium (IGM), shaping the thermal and ionization history of the early Universe.

Measurements of the 21-cm signal can be conducted using either a single-antenna radiometer or an interferometric array, each offering a different aspect of cosmological information. The single-element radiometer captures the sky-averaged 21-cm brightness temperature and reflects a part of the ionization history. Over the past decade, global 21-cm experiments have yielded a wealth of results that transforms our conventional wisdom of astrophysical mechanisms governing reionization. Observations with the Experiment to Detect the Global Epoch of Reionization Signature (EDGES) collaboration reported the first detection of an absorption profile centring at 78 MHz in the spatially averaged 21-cm spectrum at an amplitude of $500\ \mathrm{mK}$ \citep{2018Natur.555...67B}. \rev{The signal exhibited a depth that could not be explained by standard 21-cm models describing the Cosmic Dawn and was attributed to excess radio photon production rates in the early Universe} \citep{2018ApJ...868...63E,2018ApJ...858L..17F,2019MNRAS.483.5329J,2019MNRAS.483.1980M,2019MNRAS.486.1763F,2020MNRAS.499.5993R} or artefacts from the signal processing pipeline and instrumental systematic errors \citep{2018Natur.564E..32H,2019ApJ...874..153B,2019ApJ...880...26S,2020MNRAS.492...22S}. However, observations from the Shaped Antenna Measurement of Background Radio Spectrum (SARAS) did not support the finding \citep{2022NatAs...6..607S} and disfavoured models with insufficient X-ray heating from the first-generation stars \citep{2022MNRAS.513.4507B}. The recent non-detection of the SARAS 3 experiment also disfavoured radio galaxies with mass $4.4 \times 10^5 M_\odot \leq M \leq 1.1 \times 10^7 M_\odot$ at $z = 20$ and also with efficient production of radio photons \citep{2022NatAs...6.1473B}.

Interferometric observations, on the other hand, measure the fluctuations of the 21-cm signal brightness temperature field across the sky. Past and ongoing interferometric experiments have placed increasingly stringent upper constraints on the 21-cm power spectrum. The Murchison Widefield Array\footnote{\url{http://www.mwatelescope.org/}} (MWA) Phase I placed a upper limit of $\Delta_{21}^2 \leq (62.4\ \mathrm{mK})^2$ at $k = 0.2\ h\ \mathrm{cMpc}^{-1}$ and $z = 7.0$ on the 21-cm power spectrum structure with the advanced RFI mitigation techniques \citep{2019ApJ...884....1B}. \citet{2019ApJ...887..141L} thereafter reported the results for Phase II of MWA, with the best measurement at $z = 6.5$ and $k = 0.6\ h\ \mathrm{cMpc}^{-1}$ of $\Delta_{21}^2 \leq (48.9\ \mathrm{mK})^2$. The same analysis was then conducted by \citet{2023MNRAS.521.5120K} using a complementary delay
spectrum pipeline and produced a consistent result of $\Delta_{21}^2 \leq (67.7\ \mathrm{mK})^2$ at $k = 0.2\ h\ \mathrm{cMpc}^{-1}$ and $z = 7.1$. The MWA EoR1 observing field was also explored at $z = 6.5$ by \citet{2021MNRAS.508.5954R} that reported a best $2\sigma$ upper limit at $\Delta_{21}^2 \leq (73.8\ \mathrm{mK})^2$ at $k \approx 0.1\ h\ \mathrm{cMpc}^{-1}$. With the Low Frequency Array\footnote{\url{http://www.lofar.org/}} (LOFAR) High-Band Antenna (HBA), \citet{2017ApJ...838...65P} presented a best 2$\sigma$ measurement of $\Delta_{21}^2 \leq (79.6 \ \mathrm{mK})^2$ at $k = 0.053\ h\ \mathrm{cMpc}^{-1}$ in the range $z = 9.6 - 10.6$. \citet{2020MNRAS.493.1662M} subsequently analysed 141 hours of data from the LOFAR observations and improved this power spectrum constraint, with the lowest 2$\sigma$ upper limit of $\Delta_{21}^2 \leq (72.9\ \mathrm{mK})^2$ at $k = 0.053\ h\ \mathrm{cMpc}^{-1}$, ruling out the cold IGM scenarios at $z = 8.7 - 9.6$. Meanwhile, the same observation data suggest that the excess radio coefficient (relative to the CMB temperature at 78 MHz) $A_\mathrm{r} < 182$ at 95\% confidence interval, rejecting strong radio contributions from the high-redshift Universe \citep{2020MNRAS.498.4178M}. Besides, the New
Extension in Nançay Upgrading LOFAR (NenuFAR) obtain a best $2\sigma$ upper limit of $(4.9\times10^3\ \mathrm{mK})^2$ at $z = 20.3$ and $k = 0.041\ h\ \mathrm{cMpc}^{-1}$ \citep{2024A&A...681A..62M}. More recently, improved data processing push this upper constraint to $\Delta_{21}^2 < (6.8 \times 10^2\ \mathrm{mK})^2$ at $k = 0.038\ h\ \mathrm{cMpc}^{-1}$ and $\Delta_{21}^2 < (2.2 \times 10^3\ \mathrm{mK})^2$ at $k = 0.041\ h\ \mathrm{cMpc}^{-1}$ for $z = 20.3$ and $17.0$ respectively \citep{2025MNRAS.542.2785M}. Moreover, LOFAR also update its lowest $2\sigma$ upper limit at $z = 9.16$, of \rev{$(146.61\ \mathrm{mK})^2$} at $k = 0.078\ h\ \mathrm{cMpc}^{-1}$ using a single night of observations for the 3C 196 field \citep{2025MNRAS.544.1255C}. \citet{2025A&A...698A.186M} also report the improved upper limits for the North Celestial Pole (NCP): \rev{$\Delta_{21}^2 < (68.7\ \mathrm{mK})^2$ at $k = 0.076\ h\ \mathrm{cMpc}^{-1}$, $\Delta_{21}^2 < (54.3\ \mathrm{mK})^2$ at $k = 0.076\ h\ \mathrm{cMpc}^{-1}$ and $\Delta_{21}^2 < (65.5\ \mathrm{mK})^2$} at $k = 0.083\ h\ \mathrm{cMpc}^{-1}$, for $z = 10.1, 9.1$ and $8.3$ respectively. These building constraints are expected to push detection limits for the weak 21-cm spectral transition, setting the stage for the next-generation radio interferometers, such as the Hydrogen Epoch of Reionization Array\footnote{\url{http://reionization.org/}} (HERA) \citep{2017PASP..129d5001D} and the Square Kilometre Array\footnote{\url{http://www.skatelescope.org/}} (SKA) \citep{2015aska.confE...1K}. With the recent release of the first image from one complete station ``S-8''\footnote{Details of the image release can be found from \url{https://www.skao.int/en/news/560/first-image-released-one-ska-low-station}.}, the SKA1-Low is also paving the way for direct imaging of structures from the EoR, in addition to providing statistical detection of the cosmological signal down to arcminute scales \citep{2015aska.confE..10M}. \rev{In addition, HERA Phase II report the latest $2\sigma$ upper limits on the 21 cm power spectrum, with $\Delta_{21}^2 \leq (1.06 \times 10^3 \ \mathrm{mK})^2$ at $k = 0.55\ h\ \mathrm{cMpc}^{-1}$ and $z = 16.78$ and of $\Delta_{21}^2 \leq (42.2 \ \mathrm{mK})^2$ at $k = 0.70\ h\ \mathrm{cMpc}^{-1}$ and $z = 7.05$ \citep{2026ApJ...998...33A}}

However, a major challenge the 21-cm experiments face is the presence of astrophysical foregrounds. These radio signals are several orders of magnitude brighter than the 21-cm signal. The dominant contribution of the foreground contaminants arises from Galactic synchrotron emission, primarily produced by supernovae remnants and free electrons interacting with the Galactic magnetic field. In addition, Galactic free-free emission and extragalactic radio sources further contribute to the contamination, complicating the detection of the faint cosmological signal \citep{2016MNRAS.458.2928C,2021MNRAS.500.2264H}.

Fortunately, these foregrounds exhibit greater spectral smoothness compared to the 21-cm signal. Benefiting from this intrinsic feature, \rev{efforts have been devoted to recovering the cosmological signal from these foregrounds and have been traditionally categorised as foreground avoidance and foreground removal} \citep{2015aska.confE...5C}. Avoidance techniques identify and filter the Fourier modes in which foregrounds dominate. This leaves a region for the 21-cm signal in cylindrically-averaged power spectrum, known as the EoR window. \rev{By leveraging the spectrally correlated nature of foregrounds, this technique} is employed to minimise both the bias of sky signal subtraction and impact of foregrounds \citep{2016MNRAS.458.2928C,2018ApJ...864..131K}. Foreground removal aims to subtract foreground emissions from total sky signal and consists of parametric methods, such as polynomial fitting \citep{2006ApJ...648..767M,2008MNRAS.389.1319J}, and non-parametric methods, such as Principal Component Analysis (PCA) \citep{2021MNRAS.504..208C}. 

Recent years have also seen increased interest in Bayesian methods for addressing the challenging signal separation problem in 21-cm cosmology. Various studies have proposed Bayesian modelling approaches that effectively estimate antenna gain parameters \citep{2021MNRAS.503.2457B,2022MNRAS.517..910S,2022MNRAS.517..935S}, direction-dependent antenna beam parameter \citep{2024RASTI...3..400W,2025RASTI...4...42W}, cross-coupling and reflection systematics \citep{2024MNRAS.534.2653M}, foreground and 21-cm signal parameters \citep{2016MNRAS.462.3069S,2019MNRAS.484.4152S,2019MNRAS.488.2904S,2023ApJS..266...23K,2024MNRAS.535..793B}, as well as constraining antenna primary beam\footnote{\rev{We will use the term ``beam'' to denote the primary beam of the telescope in the following text, unless otherwise specified.}} structures \citep{2025MNRAS.541..687K} and low-level systematics \citep{2025A&A...704A.205M}, from interferometric datasets.

In addition, \citet{2018MNRAS.478.3640M} introduced a non-parametric GPR approach used in a Bayesian framework, demonstrating manageable risk and improved sensitivity compared to a foreground avoidance approach in recovering the 21-cm signal from simulated LOFAR data. The GPR approach was subsequently built into a pipeline to process the LOFAR interferometric data \citep{2020MNRAS.493.1662M} to yield upper limits on the 21-cm power spectrum. \citet{2024MNRAS.527.3517M} developed this approach by replacing the original analytic kernel with a variational autoencoder (VAE) neural network that can learn the \rev{shapes of 21-cm signal power spectra} in different reionization scenarios. The VAE-based GPR is then applied to the LOFAR and NenuFAR observations to improve the upper constraints \citep{2024MNRAS.527.7835A,2024MNRAS.534L..30A,2024A&A...681A..62M,2025MNRAS.544.1255C,2025A&A...698A.186M,2025MNRAS.542.2785M}. Building on this advance in interferometric measurements, we apply the GPR technique to our simulated data and forecast its performance and limitations in future SKA-Low observations.

This paper is structured as follows. In Section~\ref{sec:simulation}, we introduce the setup of our SKA\footnote{In this work, ``SKA'' refers specifically to ``SKA1-Low'' unless explicitly indicated.} simulations and describe the telescope model and sky models used to mimic realistic interferometric observations. Section~\ref{sec:data processing} discusses the data analysis pipeline we use to process the visibilities generated from OSKAR simulations into images and power spectra. Meanwhile, we also discuss how different image-weighting schemes are utilised for source subtraction and GPR foreground mitigation. The mathematical concepts of GPR and its VAE-based extension are reviewed in Section~\ref{sec:foreground mitigation}, along with their applications in extracting 21-cm signal and performance in recent interferometric observations. We further describe the generation of the training dataset for the VAE kernel function. In addition, we discuss the methods using nested sampling to estimate the posterior distribution and global evidence for five candidate GPR models. In Section~\ref{sec:results}, we demonstrate the effectiveness of Bayesian GPR models in describing the radio sky signals and extracting the 21-cm signal. In addition, we present the first application of the Bayesian-evidence null tests to interferometric data. Finally, we summarise the signal processing pipeline and conclude on the GPR models with the model predictivity and the least biased signal recovery in Section~\ref{sec:conclusion}.

\section{Simulation} \label{sec:simulation}
The SKA simulation pipeline used in this work is undertaken by using the OSKAR simulation software \citep{dulwich_2020_3758491}. The radio simulator mocks interferometric observations by incorporating antenna layouts and sky models. This process also accounts for the time-varying beam response of the SKA telescope to ensure a realistic observation and generates visibility data at the selected timepoints. The simulations use an end-to-end pipeline adapted from \citep{2024MNRAS.533.2876O}. The sky models are modified from the sky models used in SKA Science Data Challenge 3a \citep{2025MNRAS.543.1092B}. The tracking observations are centred on a field at right ascension $\alpha = 0^\mathrm{h}$ and declination $\delta = -30\degree$. Spanning 4 consecutive hours, the hour angle drifts from -2$^\mathrm{h}$ to 2$^\mathrm{h}$. The sky signals are simulated from 122 to 134 MHz, with a step of 0.1 MHz. Only the Stokes I polarization is simulated for these signals to reduce the computing time. We therefore will not discuss the potential effects of polarization leakage on the cosmological signal in this work. Each observation is integrating over 10 seconds and this yields 1440 time steps during the entire simulated observation. We also introduce thermal noise equivalent to 1000-hour observation into the data to assume a scenario where the noise level is sufficiently suppressed for 21-cm signal detection. Here we provide a detailed description of the telescope layout and the sky models used in these simulations.

\subsection{Telescope model}

\begin{figure*}
    \centering
    \includegraphics[width=\linewidth]{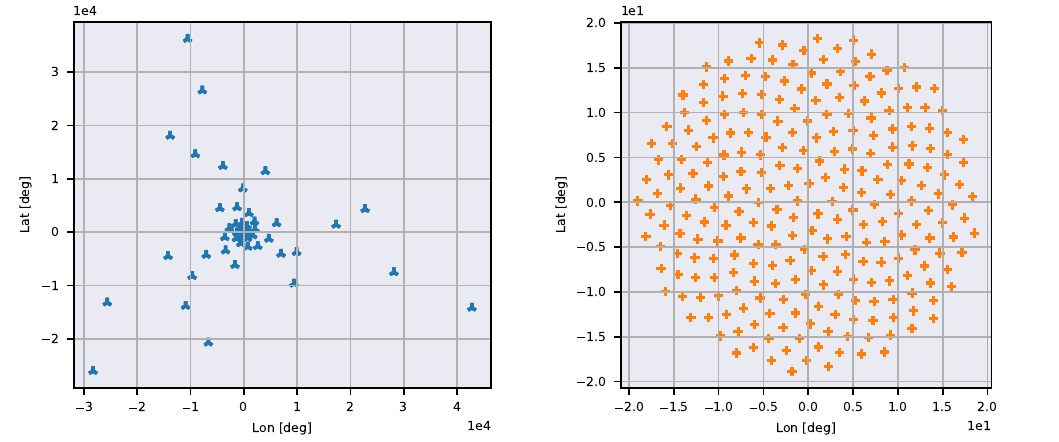}
    \caption{The position layout is shown for the telescope model used to mock SKA observations, based on \citet{Dewdney2016}. \textit{Left panel}: The geographical locations in longitude and latitude are shown for the expected SKA1-Low configuration of the 512 stations \citep{2017MNRAS.465.3680M}. The stations are arranged in a ``Vogel'' layout. \textit{Right panel}: The relative positions are shown for the 256 antenna elements within a single station. Each station is treated as identical and the beam response is simulated for a single station and duplicated for all the 512 stations.} \label{fig:ska_station_layout}
\end{figure*}

The SKA will consist of a total of 131072 log-periodic dipole antennas distributed across 512 aperture array stations as shown in Figure~\ref{fig:ska_station_layout}. The stations follow the ``Vogel'' layout, with a single-arm spiral configuration to maintain a uniform area density of antennas while maximizing azimuthal sampling.

In practice, we adopt the telescope layout based on the model proposed in \citet{Dewdney2016} and incorporate the computational optimization described in \citet{2025MNRAS.543.1092B,Bonaldi2026SKA}. Specifically, a single station layout is duplicated to all 512 stations by assuming identical configuration of antenna elements within each cluster. In this work, we model the antenna element as a dipole, restricted here to the isolated element pattern (IEP) and ignoring embedded effects. Figure~\ref{fig:ska_station_layout} shows the configurations of the 512 stations and the 256 antenna elements within one of the stations respectively. The beam duplication simplifies the effective station beam to the auto-correlation of the single station layout. However, this beamforming results in an amplified side-lobe response, compared to that formed by using 512 diverse station layouts. To compensate for this amplification, an attenuation factor of 10$^{-3}$ is applied towards the edge of the field of view (FoV). This attenuation is also intended to mimic a partially successful source calibration performed in this region. Furthermore, images that are subsequently processed confine the FoV to where the \rev{beam} is the most sensitive, further reducing the impact of side-lobes on sky signals.

\subsection{Sky model} \label{sec:sky model}
Similar to \citet{2025MNRAS.543.1092B,Bonaldi2026SKA}, the sky model is defined in two separate regions: an inner sky and an outer sky patch. The inner sky patch corresponds to the most beam-sensitive region, defined by the first null of the station beam, approximated by an \rev{Airy} beam:
\begin{eqnarray} \label{eq:inner_sky_patch}
    \rev{\theta_\mathrm{FoV} = \arcsin(\frac{1.12 \lambda}{D}),}
\end{eqnarray}
\rev{where $\lambda$ is the observing wavelength and $D$ is the length of baseline. For simplicity, the first null of the beam is determined by the longest wavelength in the bandwidth and the shortest baseline of the telescope.} The outer sky model, on the other hand, covers the entire $2\pi$ steradians above the horizon outside the inner sky patch. Meanwhile, the inner sky subtends $4.16 \degree$ at the lowest frequency channel. Since the beam response does not vary significant across the frequency bandwidth, we assumed a fixed boundary between the inner and outer sky patches over the full bandwidth for simplicity in our simulations.

\begin{figure*}
    \centering
    \includegraphics[width=\linewidth]{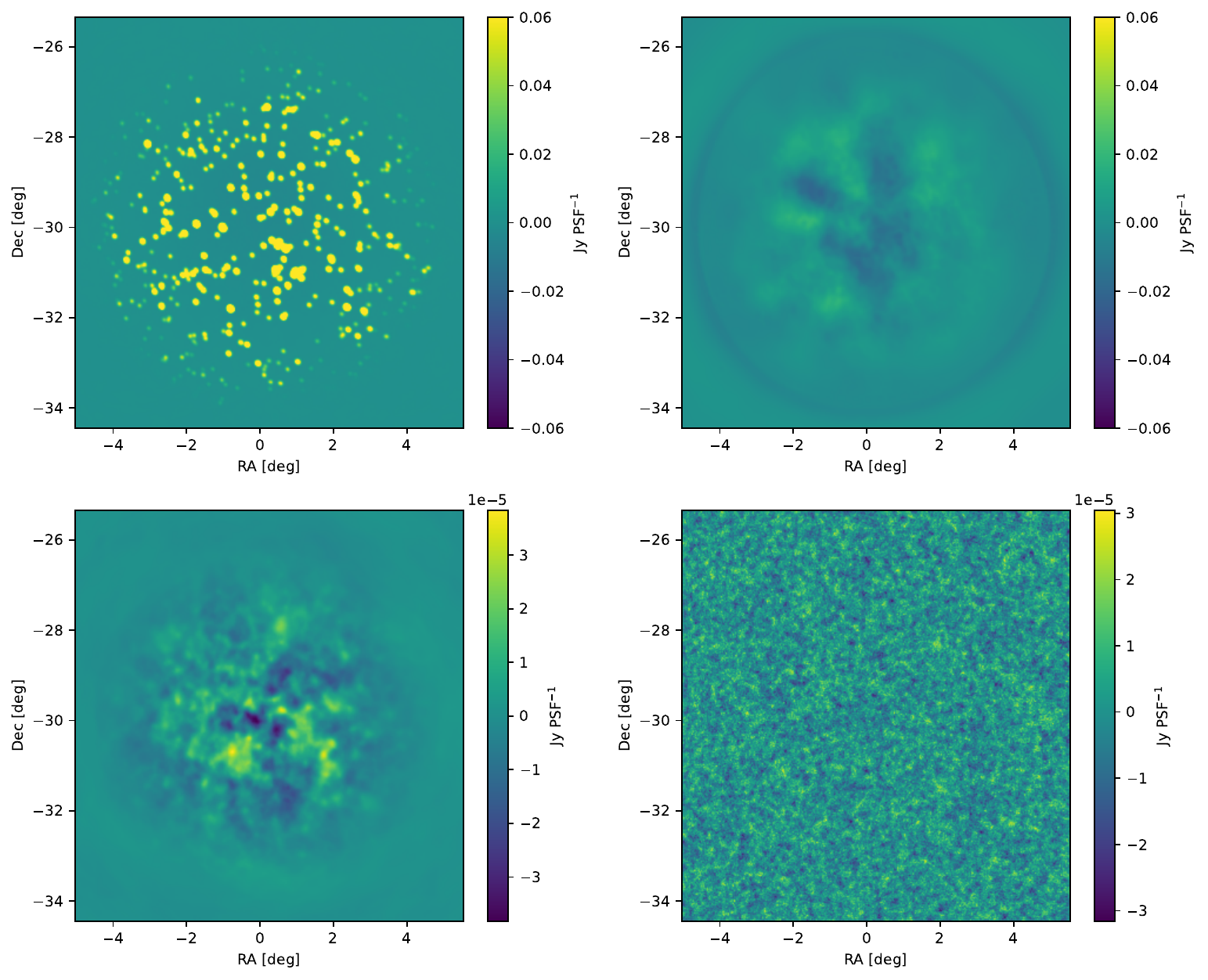}
    \caption{Sky and instrumental components simulated in Section~\ref{sec:sky model} and Section~\ref{sec:noise model} at the lowest frequency of 122 MHz. The signals are imaged using a natural deconvolved scheme, spanning a FoV of $9.1\degree$ at an resolution of 32 arcsec. \textit{Upper left}: Discrete radio sources from the GLEAM survey. The dirty beam is deconvolved from the image, which restores the source components with a Gaussian-shaped beam. The sources that fall outside the \rev{beam} as defined in Equation~\ref{eq:inner_sky_patch} are attenuated in flux density to mimic the demixing process and to balance the side-lobe amplitude from beam duplication. \textit{Upper right}: Diffuse Galactic radio emission from the GSM. The visible ring surrounding the sources marks the boundary between the outer and inner sky models. \textit{Lower left}: 21-cm brightness temperature fluctuations generated from 21cmFAST, based on astrophysical parameters listed in Table~\ref{sec:sky model}. \textit{Lower right}: Thermal noise of the interferometric array based on the telescope layout in Figure~\ref{fig:ska_station_layout}. The noise is randomly sampled from a Gaussian distribution with standard deviation determined by instrumental sensitivity from \citet{2019arXiv191212699B}. The color scales are standardised by row to facilitate comparison: the foreground emissions share the same dynamic range, while the 21-cm signal and thermal noise are displayed on a consistent fainter scale.} \label{fig:sky_grid}
\end{figure*}

The outer sky model encompasses discrete radio sources from the Galactic and Extra-Galactic All-Sky MWA Survey (GLEAM) catalogue \citep{2015PASA...32...25W,2017MNRAS.464.1146H}, as shown in the upper left panel of Figure~\ref{fig:sky_grid}. Despite far side-lobe attenuation, these sources still contribute substantially to the total sky flux density. Their flux densities are modelled as power-law functions of frequency, with their reference spectral indices at 200 MHz. Sources with invalid spectral indices are also discarded from the sky model. Diffuse emission and the 21-cm signal are not simulated in this region because the attenuation reduces their amplitudes to negligible levels, and also to save computational expense.

In the inner sky model, both diffuse radio sources and the 21-cm signal are included, in addition to the GLEAM sources described above. The diffuse foregrounds are simulated by a python package PyGDSM based on the Global Sky Model (GSM) \citep{2017MNRAS.464.3486Z}. The diffuse sky model incorporates Galactic synchrotron, free-free and dust emission. The flux density of the diffuse sources is approximated as a power-law function of frequency, with the spectral index varying spatially across the sky. A ring is visible in the upper right panel of Figure~\ref{fig:sky_grid} and marks the edge of the inner sky. Ripples are also seen beyond the ring due to the Fourier transform of finite and non-periodic sky data.

The 21-cm signal is generated using 21cmFAST \citep{2011MNRAS.411..955M,2020JOSS....5.2582M}, based on the excursion set formalism \citep{1991ApJ...379..440B}, as shown in the lower left panel of Figure~\ref{fig:sky_grid}. The semi-analytic code treats complex astrophysical processes happening during reionization with approximations and efficiently computes the brightness temperature field for the cosmological signal. The simulation initialises conditions within a box of $1536^3$ voxels. The subsequent evolution of density and ionization states is mapped onto a lower resolution grid of $512^3$ voxels, corresponding to a box size of 1223 cMpc. The simulation follows the \rev{parametrization} presented in \citet{2019MNRAS.484..933P} with the reionization parameters listed in Table~\ref{tab:ref_21cm_parameters}.

\begin{table}
\renewcommand{\arraystretch}{1.25}
\centering
\caption{EoR parameters used in 21cmFAST to simulate the 21-cm signal shown in the lower left panel of Figure~\ref{fig:sky_grid}.}
\label{tab:ref_21cm_parameters}
\begin{tabularx}{\linewidth}{@{} S Y V @{}}
\toprule
\textbf{Symbol} &
\multicolumn{1}{c}{\textbf{Description}} &
\textbf{Value} \\
\midrule
$f_\ast$ &
Fraction of galactic gas in stars, measured for $10^{10}\mathrm{M_{\odot}}$ halos &
$\log_{10}(f_\ast) = -1.509$ \\

$\alpha_\ast$ &
Power-law index of the stellar gas fraction as a function of halo mass &
$\log_{10}(\alpha_\ast) = 0.496$ \\

$f_{\mathrm{esc}}$ &
Fraction of ionizing photons escaping into the IGM, measured for $10^{10}\mathrm{M_{\odot}}$ halos &
$\log_{10}(f_{\mathrm{esc}}) = -1.046$ \\

$\alpha_{\mathrm{esc}}$ &
Power-law index of the escape fraction as a function of halo mass &
$\log_{10}(\alpha_{\mathrm{esc}}) = 0.043$ \\

$M_{\mathrm{turn}}$ &
Turnover halo mass for quenching of star formation &
$\log_{10}(M_{\mathrm{turn}}) = 8.277$ \\

$t_\ast$ &
Fraction of Hubble time that defines the star-formation rate of galaxies &
$\log_{10}(t_\ast) = 0.152$ \\
\bottomrule
\end{tabularx}
\end{table}

The output physical lightcone is transformed to an observational brightness temperature cube with a FoV of $9\degree \times 9\degree$ with unit conversion from comoving distance (cMpc, cMpc, cMpc) to local observation coordinates in sky angular scales and frequency (deg, deg, MHz). The coordinate transformation utilises the \textsc{tools21cm} Python package\footnote{\url{https://github.com/sambit-giri/tools21cm}} \citep{2020JOSS....5.2363G}. For the simulations in this work we use the $\Lambda$CDM cosmology with the Planck 2018 parameter set \citep{2020A&A...641A...6P}: $h = 0.68, \Omega_\mathrm{m} = 0.31, \Omega_\Lambda = 0.69$.

\subsection{Noise model} \label{sec:noise model}
Thermal noise originates from the random thermal motion of charge carriers within electrical conductors. It is inherently present in electrical circuits and imposes fundamental limits on the sensitivity of radio receivers. Such noise in an interferometric observation is related to the sensitivity of the telescope, and therefore the system temperature of the instrument. Instabilities of amplifiers and mixers or as a result of natural radiation, are not considered as noise here but as instrumental systematic effects instead.

% check the equation with the one in the OSKAR simulation script: per polarization
The antenna receiver temperature of SKA-Low is assumed uncorrelated across different baselines and frequencies. The system noise can therefore be treated as Gaussian distributions with zero mean and variance determined by the temperature. The root-mean-square (\textit{RMS}) thermal noise is then defined by
\begin{eqnarray} \label{eq:noise}
    N = \frac{1}{\eta_\mathrm{s}}\frac{T_\mathrm{sys}}{A_\mathrm{eff}}\frac{\sqrt{2}k_\mathrm{B}}{\sqrt{\Delta\nu \ t_\mathrm{int}}},
\end{eqnarray}
where $\eta_\mathrm{s}$ is the system efficiency coefficient, $T_\mathrm{sys}$ is the system temperature, $A_\mathrm{eff}$ is the antenna effective area, $k_\mathrm{B}$ is the Boltzmann constant, $\Delta\nu$ is the frequency bandwidth and $t_\mathrm{int}$ is the integration time. The factor $\sqrt{2}$ in Equation~\ref{eq:noise} arises from the fact that each visibility is obtained from the correlation of signals between two stations. The ratio between $T_\mathrm{sys}$ and $A_\mathrm{eff}$ provides a measure of the system sensitivity for the telescope \citep{1999ASPC..180.....T}. The natural sensitivities of the SKA1-Low array in terms of $A_\mathrm{eff}/T_\mathrm{sys}$ are listed in \citet{2019arXiv191212699B} and are interpolated using a cubic spline across the low-frequency radio band.

Equation~\ref{eq:noise} also indicates the difference between optical telescopes and radio interferometers of which the system sensitivity is independent of source sizes but is related to the observation bandwidth and integration time. Through long time exposure, thermal noise can be suppressed to a level approaching the detection limit of the cosmological signal. In our SKA simulations, thermal noise is generated by using OSKAR and is added to the sky signals, matching the expected system sensitivity for the 1000-hour integration time (see the lower right panel of Figure~\ref{fig:sky_grid}).

\section{Data processing} \label{sec:data processing}
\subsection{Imaging} \label{sec:imaging}
From the OSKAR-simulated sky data, we use the imaging software WSCLEAN to taper, grid and Fourier transform the visibilities. The software can also perform deconvolution of the dirty beam to produce cleaned images. Specifically, the imager processes ungridded visibilities, stored in CASA MeasurementSet (MS) format \citep{2022PASP..134k4501C} and described by
\begin{eqnarray} \label{eq:vis}
    \resizebox{0.9\columnwidth}{!}{$V(u,v,w) = \iint \frac{A(l,m)I(l,m)}{\sqrt{1-l^2-m^2}}e^{-2\pi i(u l+v m+w(\sqrt{1-l^2-m^2}-1))} \mathrm{d}l \mathrm{d}m,$}
\end{eqnarray}
where $u$, $v$ and $w$ are the baseline coordinates in the interferometric array, $A(l,m)$ is the \rev{beam} attenuation, $I$ is the sky specific intensity and $l$ and $m$ are cosine sky coordinates \citep{2017isra.book.....T}.

Images produced by WSCLEAN are typically convolved with the synthesised beam, also referred to as the point-spread function (PSF), determined by baseline configuration of the array and instrumental weighting. The dirty synthesised beam will be replaced with a clean beam when finer source structures need to be revealed. Meanwhile, power-spectrum estimation also requires removing the PSF dependence to convert flux density to brightness temperature. These procedures are detailed below and in Section~\ref{sec:power_spectrum_estimation}. Formally, a sky image and its PSF response image can be expressed as:
\begin{eqnarray}
    I(l,m,\nu) & = & \int V(u,v,\nu) W(u,v,\nu) e^{2\pi i(ul+vm)} \mathrm{d}u \mathrm{d}v, \label{eq:flux density} \\
    I_{\mathrm{PSF}}(l,m,\nu) & = & \int W(u,v,\nu) e^{2\pi i(ul+vm)} \mathrm{d}u \mathrm{d}v, \label{eq:psf}
\end{eqnarray}
where $V$ is the reconstructed visibility from $w$-term correction and gridding and $W$ is the weighted transfer function determined from the antenna response pattern and any applied weighting \citep{2017isra.book.....T}. In the context of wide-field imaging, the approximation that $\sqrt{1-l^2-m^2} = 1$ in Equation~\ref{eq:vis} does not hold and sky curvature needs to be taken into account. In this case, a ``w-term'' is used to characterise deviation of the telescope array from an ideal planar configuration \citep{2014MNRAS.444..606O}.

Based on the w-terms found in the data, raw visibilities are sampled into 1000 $w$-layers to account for the sky curvature. The sampled data are mapped onto a regular grid using a Kaiser-Bessel convolution kernel to optimise the gridding efficiency without compromising accuracy \citep{2014MNRAS.444..606O,2019A&A...631A..12O}. The kernel estimates the contributions to each cell from neighboring samples on the $uv$-grid using the Kaiser-Bessel window function. The kernel size and oversampling factor are set to 15 and 4095 respectively to confine the systematics significantly below the 21-cm signal. 

The exact gridding process is also determined by the image weighting scheme \citep{2017isra.book.....T}. In this data processing pipeline, we generate three types of images that serve different purposes. The images can be generally gridded under either natural or uniform weighting schemes. Natural weighting assigns constant weights to visibilities before gridding. Images generated under this scheme are maximised in sensitivity. However these images are also dominated by baselines with the highest sampling density in $uv$-space and often have poorly shaped \rev{PSF} \citep{2017isra.book.....T}. We use the CLEAN algorithm to deconvolve the dirty \rev{PSF} by iterating over the highest intensity points in the images and subtracting the response to a point source. The sources are subsequently convolved with a Gaussian beam response to yield clean-beam images. The joined-channel deconvolution in WSCLEAN extends this algorithm and further allows us to clean the images on multiple size scales \citep{1974A&AS...15..417H,4703304,2017MNRAS.471..301O}.
With this multi-scale cleaning technique, we can also identify potential diffuse radio sources in the observed data in addition to point sources.

In contrast to natural weighting mode, uniform weighting gives weights inversely proportional to the local density in the $uv$ plane. The density depends on the number of visibility data points within a defined radius. This ensures contributions from different spatial scales are more consistently weighted to minimise the side-lobe levels. Nevertheless, the weighting process may amplify small-scale structures and result in reduced sensitivity. To minimise the impact of the \rev{PSF} on individual radio sources, we also apply beam deconvolution to the uniformly-weighted images. These images are therefore suitable for point source calibration because of the cleaned beam response and the effectively suppressed diffuse structures. We refer to the deconvolved, naturally-weighted images as ``natural images'' and the deconvolved, uniformly-weighted images as ``uniform images'' in the following context unless otherwise indicated.

To balance image resolution and computational efficiency, we configure the image dimensions to 1024 $\times$ 1024 pixels with a resolution of 32 arcsec for natural images. The image size and resolution are set to 2048 pixels and 16 arcsec for uniform images. This higher spatial resolution facilitate the identification of small-scale structures (i.e. the point sources) in the images. Table~\ref{tab:imaging} provides a detailed overview of the imaging parameters we use in WSCLEAN. We also employ an identical set of deconvolution parameters for both imaging weighting modes: $\mathrm{niter} = 1000000$, $\mathrm{mgain} = 0.8$, $\mathrm{auto\ mask} = 4$ and $\mathrm{auto\ threshold} = 1$.

\begin{table}
\renewcommand{\arraystretch}{1.2}
\centering
\caption{Configuration used in WSCLEAN to generate naturally and uniformly weighted images from OSKAR-simulated visibilities. Natural images are used for GPR foreground mitigation due to their higher sensitivity. Higher resolution is applied to uniform images for better point source detection. Images in both weighting schemes cover the same FoV of 9.1\degree. The images are also deconvolved using the parameters listed separately in Section~\ref{sec:imaging}.} \label{tab:imaging}
\begin{tabular}{@{}ccc@{}}
\toprule
\textbf{Configuration} & \textbf{GPR removal} & \textbf{Source detection} \\
\midrule
weight & natural & uniform \\
scale & 1024 & 2048 \\
resolution & 32 asec & 16 asec \\
taper-gaussian & 60 & 60 \\
taper-edge & 100 & 100 \\
padding & 2 & 2 \\
wstack-nwlayers & 1000 & 1000 \\
wstack-oversampling & 4095 & 4095 \\
wstack-grid-mode & kb & kb \\
wstack-kernel-size & 15 & 15 \\
\bottomrule
\end{tabular}
\end{table}

\subsection{Radio source subtraction}
In this section, we perform a preliminary foreground subtraction to suppress the total foreground flux density. These procedures aim to provide a cleaner input for the GPR-based signal extraction and reduce the bias in the 21-cm signal recovery. We therefore focus on identifying and subtracting bright point sources while attenuating diffuse foregrounds in the observed sky signal. The pre-processing procedures discussed in this section are based on the source subtraction methods described in \citet{mertens_2023_10263162} and \citet{2025MNRAS.543.1092B,Bonaldi2026SKA}.

% In practice, we find GPR is not a well-suited calibrator for accurately calibrating sources scattered sparsely across the sky, i.e. point sources. To address this limitation,

\subsubsection{Compact radio sources} \label{sec:compact radio sources}
We identify and extract compact point sources using the PyBDSF package\footnote{\url{https://pybdsf.readthedocs.io/en/latest/}}. The source-finding algorithm calculates local background \textit{RMS} noise across the image while a threshold is set to differentiate islands of radio sources and noise. Islands are further decomposed into Gaussian components that \rev{characterise} the flux, size, and position angle of detected sources.

The source detection is employed using uniform images from Section~\ref{sec:imaging} in which individual sources are clearly identified. Point source calibration often requires imaging visibility data using long baselines to better account for the small scales of point sources. However, in our data we find that uniform weighting effectively attenuates extended structures, making baseline filtering unnecessary. Furthermore, we notice that discarding short baselines introduces artefacts that could obscure source identification. Therefore, we retain the full range of spatial scales in this step.

\rev{To leverage the $uv$-coverage and minimise the noise level in the data, we perform source detection on images derived from the full 1440-timestep observations and
find a promising detection with negligible w-layer eﬀect in our simulation. The w-layer effect
originates from images synthesised from the 1440 time samples that span a large area of the
curved sky. However, this effect is minimised by the w-stacking during imaging.} Meanwhile, the detected source flux density is also subject to attenuation by the \rev{beam} that varies spectrally and spatially. This variation introduces biases in flux estimation and therefore spectral index prediction. To correct for this underestimation of source flux, we simulate the \rev{beam} response using the \textsc{oskar\_sim\_beam\_pattern} implementation in OSKAR, using the telescope configuration as in our visibility simulation. \rev{The beam response is also computed across the same frequency range and is averaged over four hours, consistent with the integration time of the full observations.} The image size and resolution are also made consistent with those of the uniform images as described in Section~\ref{sec:simulation} and Section~\ref{sec:imaging}. The unattenuated flux density of each source is estimated by applying a correction factor using the relationship, $I_{\mathrm{corr}} = I_{\mathrm{apparent}}/B$, where $I_{\mathrm{corr}}$ is the beam corrected flux density,  $I_{\mathrm{apparent}}$ the apparent flux density observed from images and $B$ the simulated \rev{beam} response. The beam-corrected images are then processed with PyBDSF, using the source-fitting parameters listed in Table~\ref{tab:source fitting}.

\begin{table}
\renewcommand{\arraystretch}{1.2}
\centering
\caption{Source detection parameters used in PyBDSF. The ``rms\_box'' parameter defines the region over which background \textit{RMS} noise is estimated. Source islands are identified based on their flux density distribution from the local background using the ``threshold\_isl'' parameter. Individual sources within these islands are flagged based on the ``threshold\_pix'' criterion.} \label{tab:source fitting}
\begin{tabular}{ccc}
\toprule
\textbf{Detection parameter} & \textbf{Value} & \textbf{Unit} \\
\midrule
rms\_box & (30,10) & pixel \\
threshold\_isl & 20 & sigma \\
threshold\_pix & 15 & sigma \\
\bottomrule
\end{tabular}
\end{table}

In PyBDSF, an ``rms box'' defines the region in which the background noise map, is calculated. We set the box size to 30 pixels with a step of 10 pixels. A region is classified as a fitting island when the local flux exceeds 20 sigma above the local mean flux, as determined by threshold\_isl. Moreover, threshold\_pix sets the source detection threshold such that a source is flagged when the local flux is 15 sigma above the mean flux.

Following the source extraction process, we obtain a catalogue of compact sources identified from the observed data. We further verify the reliability of detections by examining the fitted spectral indices. The power-law indices are expected to range from -2.5 and 2.7. This range is observed from the GLEAM sources used in our simulation, but is adjusted with a slight extension to account for potential errors between the model and observed flux. We also flagged the sources with spectral indices of zero as misidentifications since the sky model does not encompass sources with uniform spectral flux density.

\subsubsection{Extended radio emission} \label{sec:extended radio emission}
Pre-cleaning of the diffuse sources is conducted by using the ``multi-scale deconvolution algorithm in WSCLEAN \citep{2017MNRAS.471..301O}. This technique models radio structures as being composed of various different scales. \citet{2022A&A...662A..97G} uses a Braggs +0.5 weighting scheme to balance sensitivity and shape of the \rev{PSF}. Our pipeline, however, employs the natural images from Section~\ref{sec:imaging} to maximise the sensitivity. This is found to preserve more diffuse structures after cleaning and is advantageous for modelling the GSM sources from the images.

The deconvolution is also performed across the full bandwidth to reduce the risk of both mislocating the sources and misidentifying noise peaks or side-lobes as radio sources \citep{2017MNRAS.471..301O}. In terms of the spectrally correlated nature of foregrounds, the joined-channel deconvolution across the full bandwidth from 122 to 134 MHz also ensures spectral continuity of detected sources and helps avoid inadvertently fitting the 21-cm signal. This cleaning process generates a catalogue of diffuse foreground sources. Each source is modelled with a Gaussian representation that allows OSKAR to reconstruct their visibilities.

\subsubsection{Sky model reconstruction}
Reconstructed from the sky models obtained in Section~\ref{sec:compact radio sources} and Section~\ref{sec:extended radio emission}, we predict the visibilities of both compact and extended sources using OSKAR. The simulation settings are kept consistent with those described in Section~\ref{sec:simulation}. Point sources are simulated according to the source position and flux density provided to OSKAR. Extended sources, in addition to their location and brightness, are further specified by the major and minor axes in unit of the full width at half maximum (FWHM) and the position angle of the major axis. In both cases, the source flux density is estimated by a power law with the spectral index and the reference flux obtained from the steps above.

\begin{figure}
    \centering
    \includegraphics[width=\linewidth]{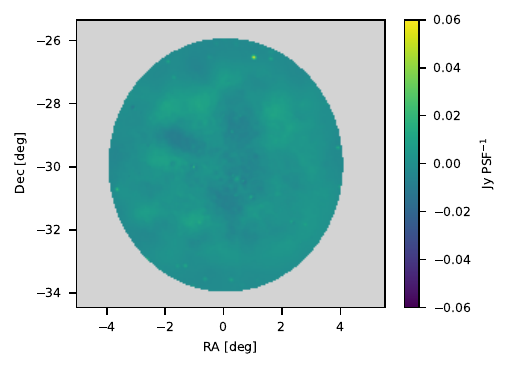}
    \caption{Deconvolved source-subtracted radio sky, in the same frequency range and imaging resolution as in Figure~\ref{fig:sky_grid}. However, the FoV is restricted to the central $4\degree \times 4\degree$ region due to the maximised beam sensitivity for the following power spectral analysis. The residual sky is obtained by calibrating and subtracting the astrophysical foregrounds described in Section~\ref{sec:compact radio sources} and Section~\ref{sec:extended radio emission} from the total sky signal. The faint residual sources in the image are due to the potential calibration errors, including beam chromaticity and fitting errors. These residuals are further mitigated using the Gaussian Process Regression (GPR) models described in Section~\ref{sec:foreground mitigation}.} \label{fig:sky_residual_natural_deconv_image}
\end{figure}

A pre-cleaned sky signal is obtained by subtracting these reconstructed foreground models from the observed data. The computation is performed directly in ungridded visibilities to preserve the most information from observations. Residual visibilities are therefore expected to contain a significantly lower level of foreground emissions. We subsequently conduct the same imaging process as described in Section~\ref{sec:imaging} to generate source-cleaned images for the following analysis (see Figure~\ref{fig:sky_residual_natural_deconv_image}).

\subsection{Power spectrum estimation} \label{sec:power_spectrum_estimation}

An image-based power spectrum is the Fourier representation of sky images in power units. The Fourier transform along the line of sight is taken at a fixed angular scale to achieve true $k_\perp-k_\parallel$ orthogonality. This differs from delay spectra obtained from direct frequency transform of ungridded visibilities where each delay is tied to a fixed physical baseline and thus couples to transverse structures \citep{2019MNRAS.483.2207M}. To conduct the image power spectral analysis, the  ps\_eor\footnote{\url{https://gitlab.com/flomertens/ps_eor}} Python package is utilised to process the images to gridded visibilities before estimating their power spectra. 

Here we explain the procedures involved in power spectral measurements. Since the three-dimensional Fourier power spectrum is defined as
\begin{eqnarray}
    P(\mathbf{k}) & = & \mathbb{V_\mathrm{c}}|\Tilde{T}(\mathbf{k})|^2, \\
    \Tilde{T}(\mathbf{k}) & = & \frac{1}{N_l N_m N_\nu} \sum_\mathbf{r} T(\mathbf{r})e^{-2i\pi\mathbf{kr}}, \label{eq:dft} \\
    \mathbb{V}_\mathrm{c} & = & \frac{L_l L_m L_\nu}{\braket{A(l,m)^2 W_s(l,m)^2} \braket{W_\nu(\nu)^2}}, \label{eq:cosmic volume}
\end{eqnarray}
where $\Tilde{T}(\mathbf{k})$ is the Fourier transformed brightness temperature and $\mathbb{V_\mathrm{c}}$ is the comoving cosmological volume,
normalised by the \rev{beam} $A(l,m)$, spatial tapering function $W_s(l,m)$ and frequency tapering function $W_\nu(\nu)$. The spatial and frequency dimensions are denoted by $N_l, N_m, N_\nu$, along with the transverse comoving scale $L_i = N_i \Delta i\cdot D_M$ where $i = \{l,m\}$, the line-of-sight comoving distance $L_\nu = N_\nu \Delta \nu\cdot D_C$ \citep{2014PhRvD..90b3018L,2020MNRAS.493.1662M}.

According to Equations~\ref{eq:flux density} and \ref{eq:psf}, the unit conversion from flux density to brightness temperature requires removing the PSF dependence from sky images:
\begin{eqnarray}
    \Tilde{I}'(u,v,\nu) & = & \frac{\Tilde{I}(u,v,\nu)}{\Tilde{I}_{\mathrm{PSF}}(u,v,\nu)}, \label{eq:de-psf} \\
    \Tilde{T}(u,v,\nu) & = & \frac{10^{-26}c^2}{2k_\mathrm{B}\nu^2\Delta l \Delta m} \Tilde{I}'(u,v,\nu), \label{eq:jytok}
\end{eqnarray}
where $\Tilde{I}$ and $\Tilde{I}_\mathrm{PSF}$ denote the spatial Fourier transform of image and PSF data, $\Delta l$ and $\Delta m$ represent the pixel sizes (i.e. resolution) in the $l$ and $m$-directions.

In signal processing, the dataset is typically non-periodic and has finite length. This is equivalent to a rectangular truncation of an infinite signal, leading to Fourier mode leakage and side-lobes. To reduce such bias, window functions are necessary in both spatial and frequency transform to smooth the abrupt discontinuities at the boundaries and can therefore effectively suppress side-lobes from non-periodicity of the observed data. Additionally, the power-spectrum estimation requires converting the spatial coordinates of image cubes in angular scales to transverse comoving distance $D_\mathrm{M}$. Meanwhile, frequencies are mapped to line-of-sight comoving distance $D_\mathrm{C}$. However, for the cosmological model used in this work (see Section~\ref{sec:sky model}), $D_\mathrm{M}$ and $D_\mathrm{C}$ are equated due to the negligible curvature.

% The zero-frequency component corresponds to the mean of the dataset and is usually subtracted from the dataset to avoid mode leakage. This Fourier mode in power spectrum measurement is interpreted as the measurement at infinitely large physical scale. Since most interferometers do not sample the zero baseline\footnote{Although specific interferometric arrays can be configured to measure the monopole, here we assume that the overlapping beam response from two closely placed antennas is negligible.}, it is therefore noted that the visibility products are already mean-centred in each frequency channel.

\begin{figure*}
    \centering
    \includegraphics[width=\linewidth]{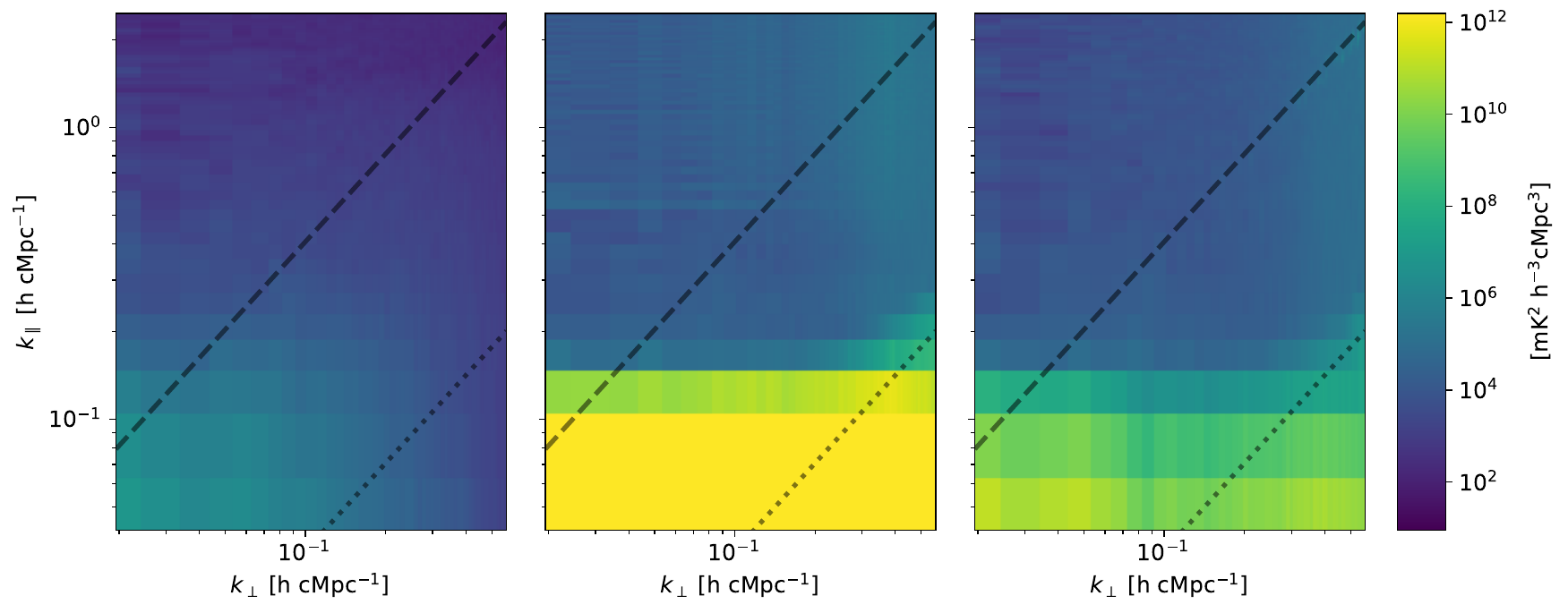}
    \caption{Cylindrically averaged 2D power spectra $P(k_\perp,k_\parallel)$ for the 21-cm signal (left panel), the full sky (middle panel) and the foreground pre-subtracted sky (right panel). The power spectra are all estimated from the central $4\degree \times 4\degree$ regions of Figure~\ref{fig:sky_grid} and Figure~\ref{fig:sky_residual_natural_deconv_image}. The dashed line indicates the horizon limit and the dotted line indicates the first null of the beam. The full sky, shown prior to any foreground subtraction, comprises discrete sources, diffuse emission, the 21-cm signal, and instrumental thermal noise. A consistent dynamic range is maintained across the three panels to ensure comparability.} \label{fig:eor_sky_ps2d}
\end{figure*}

The 21-cm signal from the EoR is generally assumed to be statistically isotropic across the sky. The physical properties are therefore expected to remain the same as observed from all directions. The isotropy also implies the statistical measurements of the signal, such as power spectrum distribution, are spherically symmetric \citep{2004ApJ...615....7M}. The measurements taken within a shell of Fourier space are statistically identical and therefore can be combined to determine the structures of the signal. In three-dimensional power spectra, wavenumber $\mathbf{k}$ can be decomposed into each dimension as,
\begin{eqnarray} \label{eq:k_modes_3d}
    k_x = \frac{2\pi u}{D_M(z)},\ 
    k_y = \frac{2\pi v}{D_M(z)},\ 
    k_z = \frac{2\pi H_0 \nu_{21} E(z)}{c(1+z)^2} \tau_g,
\end{eqnarray}
where $c$ is the light speed, $H_0$ is the Hubble constant, $\nu_{21}$ is the rest-frame frequency of the \ion{H}{I} spin-flip transition, $E(z)$ is the evolution function for the Hubble parameter and $\tau_g$ is the geometric delay. By averaging over the spatial axes $x$ and $y$, one can also conduct a cylindrical power measurement from the 3D Fourier representation of the signal (see Figure~\ref{fig:eor_sky_ps2d} for the cylindrical power spectra of the 21-cm signal and the full sky): 
\begin{eqnarray}
    k_\perp & = & \sqrt{k^2_x + k^2_y}, \label{eq:k_perp} \\
    k_\parallel & = & k_z. \label{eq:k_par}
\end{eqnarray}

Furthermore, the power spectrum can be averaged in spherical shells due to the isotropy, described by $k = \sqrt{k^2_\perp + k^2_\parallel}$. The spherically averaged dimensionless power spectrum is defined by:
\begin{eqnarray} \label{eq:dimensionless_ps1d}
    \Delta^2(k) = \frac{k^3}{2\pi^2}P(k).
\end{eqnarray}

% Figure~\ref{fig:sky_precal_ps1d} shows an example of the binned spherical 1D power spectra for the full sky and source-subtracted sky.

\section{Foreground mitigation} \label{sec:foreground mitigation}
We implement the novel VAE-based GPR approach \citep{2024MNRAS.527.3517M} on the pre-processed SKA data obtained from Section~\ref{sec:data processing}. This technique has been applied to recent radio-interferometric observations, yielding updated upper limits on the 21-cm power spectrum \citep{2024MNRAS.527.7835A,2024MNRAS.534L..30A,2024A&A...681A..62M,2025MNRAS.544.1255C,2025A&A...698A.186M,2025MNRAS.542.2785M}. Here we briefly review the fundamental principles of GPR in radio foreground mitigation, followed by applying Bayesian GPR models to the simulated SKA data.

\subsection{Gaussian Process Regression} \label{sec:gpr}
\citet{2018MNRAS.478.3640M} proposed a signal extraction approach to separate the 21-cm signal from spectrally smooth foregrounds and mode-mixing effects. Each sky component is represented by a Gaussian process (GP) model using this approach. These GPs are probabilistic models that extend the concept of finite-dimensional to function-space distributions. Each model therefore defines a distribution over all possible latent functions $f$ that fit a set of points \citep{2006gpml.book.....R}, governed by a mean function $\mu$ and a covariance function (also termed as kernel function) $\kappa$, such that
\begin{eqnarray} \label{eq:gp}
    f \sim \mathcal{GP}(\mu,\kappa).
\end{eqnarray}

For any collection of inputs $\mathbf{x} = \{\mathrm{x_1}, \ldots, \mathrm{x_n}\}$, there exist Gaussian distributed random variables $\mathbf{f} = f(\mathbf{x})$. The joint distribution of any finite set of these variables follows a multivariate Gaussian distribution defined by,
\begin{eqnarray}
    \mathbf{f} \sim \mathcal{N}(\mu(\mathbf{x}),\kappa(\mathbf{x},\mathbf{x})),
\end{eqnarray}
where the mean vector $\mu(\mathbf{x})$ provides the prior trend of $f(\mathbf{x)}$ and the kernel $\kappa(\mathbf{x},\mathbf{x})$ is a symmetric positive semidefinite matrix with entries $\kappa_{ij} = \kappa(\mathrm{x}_i,\mathrm{x}_j)$ and encodes assumptions about function behaviour. While applying GP models to regression tasks, predictions are based on the latent functions $f$ that model the observations. Given a set of new inputs $\mathbf{x'}$, the predicted function outputs are $\mathbf{f'} = f(\mathbf{x'})$ and the joint density distribution of $\mathbf{f}$ and $\mathbf{f'}$ is described by

\begin{eqnarray} \label{eq:jointdis}
    \begin{bmatrix}
        \mathbf{f} \\ \mathbf{f'}
    \end{bmatrix} \sim \mathcal{N}
    \begin{pmatrix}
        \begin{bmatrix}
            \mu(\mathbf{x}) \\ \mu(\mathbf{x'})
        \end{bmatrix} ,
        \begin{bmatrix}
            \kappa(\mathbf{x},\mathbf{x}) & \kappa(\mathbf{x},\mathbf{x'}) \\
            \kappa(\mathbf{x'},\mathbf{x}) & \kappa(\mathbf{x'},\mathbf{x'})
        \end{bmatrix}
    \end{pmatrix}.
\end{eqnarray}
Hence conditioning the joint distribution in Equation~\ref{eq:jointdis} on $\mathbf{f}$ yields the predictive distribution of $\mathbf{f'}$ for the regression tasks:
\begin{eqnarray} \label{eq:gp_conditional}
\mathbf{f'}|\mathbf{f},\mathbf{x},\mathbf{x'} \sim \mathcal{N}(\langle \mathbf{f'} \rangle,\mathrm{cov}(\mathbf{f'})),
\end{eqnarray}
where $\langle \mathbf{f'} \rangle$ and $\mathrm{cov}(\mathbf{f'})$ are the predictive mean and predictive covariance respectively \citep{2006gpml.book.....R,10360364}. The mean and covariance for the predictions at $\mathbf{x'}$ are given by,
\begin{eqnarray} \label{eq:gp_mean_cov}
    \langle \mathbf{f'} \rangle & = & \kappa(\mathbf{x'},\mathbf{x})\kappa(\mathbf{x},\mathbf{x})^{-1}\mathbf{f}, \\
    \mathrm{cov}(\mathbf{f'}) & = & \kappa(\mathbf{x'},\mathbf{x'}) - \kappa(\mathbf{x'},\mathbf{x})\kappa(\mathbf{x},\mathbf{x})^{-1}\kappa(\mathbf{x'},\mathbf{x}).
\end{eqnarray}

In the simulations we described in Section~\ref{sec:sky model}, the observed data can be described by, 
\begin{eqnarray} \label{eq:signal}
    \mathbf{d} = \mathbf{f}_{21} + \mathbf{f}_{\mathrm{fg}} + \mathbf{n},
\end{eqnarray}
The mean function is usually set to $\mu(\nu) =0$. The combined kernel function is additive and can be derived from Equation~\ref{eq:signal} as a linear combination of the composites,
\begin{eqnarray} \label{eq:kernel}
     \mathbf{K} = \mathbf{K}_\mathrm{21} + \mathbf{K}_\mathrm{fg} + \mathbf{K}_\mathrm{n}.
\end{eqnarray}
The foreground covariance can be decomposed into intrinsic and instrumentally mixed components with different coherence scales, $\mathbf{K}_\mathrm{fg} = \mathbf{K}_\mathrm{int} + \mathbf{K}_\mathrm{mix}$ \citep{2018MNRAS.478.3640M,2020MNRAS.493.1662M}. While beam chromaticity affects all sky signals, $\mathbf{K}_\mathrm{int}$ denotes the spectrally smoother, low-$k_\parallel$ component and $\mathbf{K}_\mathrm{mix}$ captures the less smooth component that populates the ``wedge'' in the cylindrical power spectrum as shown in the middle panel of Figure~\ref{fig:eor_sky_ps2d}. Additionally, the joint density distribution of the total observed signal and the function values for a sky signal component $\mathbf{f}_i$ at frequency $\nu'$ is,
\begin{eqnarray} \label{eq:eor_jointdis}
    \begin{bmatrix}
        \mathbf{d} \\ \mathbf{f}_i
    \end{bmatrix} \sim \mathcal{N}
    \begin{pmatrix}
        \begin{bmatrix}
            0 \\ 0
        \end{bmatrix} ,
        \begin{bmatrix}
            \mathrm{K}_\mathrm{fg}(\nu,\nu) + \mathrm{K}_\mathrm{21}(\nu,\nu) + \mathrm{K}_\mathrm{n}(\nu,\nu) & \mathrm{K}_i(\nu,\nu') \\
            \mathrm{K}_i(\nu',\nu) & \mathrm{K}_i(\nu',\nu')
        \end{bmatrix}
    \end{pmatrix}.
\end{eqnarray}
However, before predicting each sky component $\mathbf{f}_i$, parameters within the kernel functions need to be optimised to achieve the maximised model performance. The optimization is carried out by maximizing the log-marginal likelihood (LML) that is the likelihood function marginalised over all latent functions and is given by,
\begin{eqnarray} \label{eq:gpr lml}
    \log P(\mathbf{d}|\nu,\theta) = -\frac{1}{2} \mathbf{d}^{\mathrm{T}}\mathbf{K}^{-1}\mathbf{d} - \frac{1}{2}\log|\mathbf{K}| - \frac{N}{2}\log2\pi,
\end{eqnarray}
where $\theta$ represents the GPR model parameters and $N$ is the number of data points. The first term of Equation~\ref{eq:gpr lml} tells how well the model fits individual observed data points. This term penalises large residuals between the model predictions and actual observations and is reduced when the model well represents the data. The second term acts as a regularization term and accounts for the model complexity. Kernels that come with significant complexity are penalised for introducing excessive covariance. The last term is a constant term that ensures the LML is a properly normalised probability density function.

\subsection{VAE kernel function}
Previous GPR-based studies have shown that the Mat\'ern kernels are widely applicable for modelling the spectral behaviours of both intrinsic foregrounds and mode-mixing contaminants, as well as the 21-cm signal \citep{2018MNRAS.478.3640M,2020MNRAS.493.1662M,2020MNRAS.495.2813G}. This class of kernels is in the general form:
\begin{eqnarray} \label{eq:matern}
    \mathrm{K_{Mat\acute{e}rn}}(\nu,\nu') = \sigma^2 \frac{2^{1-\eta}}{\Gamma(\eta)} (\sqrt{2\eta}\frac{|\nu-\nu'|}{l})^\eta K_\eta (\sqrt{2\eta}\frac{|\nu-\nu'|}{l}),
\end{eqnarray}
where $\sigma^2$ describes the variance of the signal, $\eta$ is the spectral parameter of the kernel function, $l$ denotes the length-scale or spectral coherence scale, and $\Gamma$ and $K_\eta$ are the gamma function and modified Bessel function of the second kind respectively. The adaptable parameters in Equation~\ref{eq:matern} allow the Mat\'ern kernel to characterise signals with varying degrees of smoothness. By varying the spectral parameter $\eta$ from $1/2$ to infinity, the function progressively smooths out and enables it to describe radio signals with larger frequency coherence scales. Further details on the effects of different values of $\eta$ are discussed in \citet{2018MNRAS.478.3640M,2020MNRAS.493.1662M,2020MNRAS.495.2813G,2025arXiv251003742R}.

However, the 21-cm signal from the Epoch of Reionization (EoR) is explained by a set of astrophysical and cosmological parameters in most reionization models, leading to complex spatial and spectral fluctuations \citep{2006PhR...433..181F,2011MNRAS.411..955M,2016MNRAS.457.1864L,2019MNRAS.484..933P}. Standard analytic kernel functions therefore often fall short in capturing the intricate structures of the cosmological signal due to their fixed prior form of covariance. The mismatch between the true 21-cm covariance and the adopted kernel function may therefore bias the cosmological signal recovery and falsely rule out EoR models \citep{2021MNRAS.501.1463K}. Hence non-analytic kernels with more variability are necessitated to accommodate this complexity in possible reionization scenarios. 

To address these issues, \citet{2024MNRAS.527.3517M} introduced a variational autoencoder (VAE) based  kernel function to learn the covariance structures of the 21-cm signal from simulations. Here we briefly review the concept of autoencoders and their extension to variation autoencoders. An autoencoder is a neural network to learn efficient representations of data, typically for dimensionality reduction or feature capturing \citep{2006Sci...313..504H}. The algorithm usually contains an encoder and a decoder. The encoder compresses the input data into low-dimensional latent space. The decoder, on the other hand, reconstructs the original data from the compressed latent dimensions. However, a deterministic autoencoder only learns a mapping that reconstructs the training data, instead of a full generative distribution. Interpolations in the latent space are therefore not guaranteed to be meaningful and often only work smoothly near the encoded training codes.

Variational autoencoders map the input data to probability distributions instead of fixed latent vectors \citep{2013arXiv1312.6114K}. Latent vectors in the VAE are therefore sampled from these distributions to reconstruct data of more flexibility. Besides, in addition to the reconstruction loss, VAEs add a regularization term, known as the Kullback–Leibler (KL) divergence, to the loss function to reward the learnt latent distributions that are aligned closely with standard Gaussian distributions. The regularization promotes smoothness and continuity in the latent space and facilitates meaningful interpolation between latent representations. A regularization parameter $\beta$, is also introduced to balance the contribution of the KL divergence during optimization. Adjusting $\beta$ allows control over the trade-off between the reconstruction loss and the smoothness of the latent space. We refer interested readers to \citet{2024MNRAS.527.3517M} for a detailed discussion of the mathematical derivation for the VAE neural network in the context of 21-cm cosmology.

\subsection{Kernel training} \label{sec:kernel training}
Since a kernel function describes the covariance in a Gaussian process, we train the VAE using a set of covariance matrices of simulated 21-cm signals. The simulations are generated using the 21cmFAST to explore the parameter space described in Section~\ref{sec:sky model}. Each parameter set corresponds to a reionization scenario and allows the VAE to characterise different morphologies of the 21-cm signal. The astrophysical parameters we use to generate the 21-cm simulations are consistent with those described in Section~\ref{sec:sky model} and their values are listed in Table~\ref{tab:21cmfast simulation}.

\begin{table}
\centering
\caption{Astrophysical parameters used to generate EoR simulations with 21cmFAST. The listed parameter values are expressed in logarithmic scale for better sampling and representation of their dynamic range.} \label{tab:21cmfast simulation}
\begin{tabular}{cccc}
\toprule
\textbf{EoR parameter} & \textbf{Min} & \textbf{Max} & \textbf{Step} \\
\midrule
$f_*$ & -3 & 0.5 & 1 \\
$\alpha_*$ & -0.5 & 1.5 & 0.5 \\
$f_\mathrm{esc}$ & -3.0 & 0.5 & 1 \\
$\alpha_\mathrm{esc}$ & -1.0 & 1.0 & 0.5 \\
$M_\mathrm{turn}$ & 8 & 10.5 & 1 \\
$t_*$ & -2 & 0 & 3 \\
\bottomrule
\end{tabular}
\end{table}

The angular correlation is estimated from the spherically-averaged 1D power spectrum of the 21-cm signal $P_\mathrm{HI}$, using the expression \citep{2007MNRAS.378..119D}:
\begin{eqnarray} \label{eq:21-cm covariance}
    C_l(\Delta \nu, k_{\perp}) = \frac{T_b^2}{\pi D^2_{\nu}} \int_{0}^{\infty} P_{\mathrm{HI}} \left( \sqrt{k^2_{\perp} + k^2_{\parallel}} \right) \cos(k_{\parallel} D_{\nu}\Delta \nu) \mathrm{d}k_{\parallel},
\end{eqnarray}
where $T_b$ is the brightness temperature fluctuation of the 21-cm signal and $D_{\nu}$ is the comoving distance at the mean frequency of the observation cube. In this step, we train the VAE to capture the signal patterns from different reionization models. The overall amplitude of the signals is described by the variance parameter in the kernel function, which we will discuss in the following section. The spherical power spectra are therefore normalised before converting to covariance training dataset. The algorithm hyperparameters of the trained VAE model are summarised in Table~\ref{tab:vae model}.

\begin{table}
\centering
\renewcommand{\arraystretch}{1.1}
\caption{Configuration of the VAE neural network. The architecture of the VAE model is implemented from the ps\_eor package and built on PyTorch. The model is trained on the spherical power spectra of simulated 21-cm signals to learn and encode their covariance structures.} \label{tab:vae model}
\begin{tabular}{cc}
\toprule
% Fitting parameter & Value & Unit \\
Encoder hidden dimension & 20,20,20 \\
Latent dimension & 2 \\
Decoder hidden dimension & 20,20,20 \\
Batch size & 128 \\
Epoch & 20000 \\
Regularization constant & 1e$^{-3}$ \\
Learning rate & 5e$^{-3}$ \\
optimiser & Adagrad \\
\bottomrule
\end{tabular}
\end{table}

\subsection{Candidate GPR models} \label{sec:candidate_gpr_models}
% provide a Bayesian framework to compare the models 
Despite the extensive application in radio foreground modelling and mitigation \citep{2020MNRAS.493.1662M,2020MNRAS.495.2813G,2021MNRAS.500.2264H,2022MNRAS.510.5872S,2024MNRAS.527.7835A,2024MNRAS.534L..30A}, direct comparisons are generally not applicable to GPR models from different studies because of different parametrizations tailored to individual observation dataset. Besides, adjusting a single model parameter will usually lead to unpredictable changes in the latent functions generated by the Gaussian processes due to the non-parametric nature. This leads to difficulty in determining the optimal \rev{parametrization} to model the target sky region. In this section, we apply five GPR models of different parametrizations to analyzing the SKA observations described in Section~\ref{sec:simulation}. 

We construct five GPR models: \mwedge{}, \anoise{}, \avarl{}, \manalytic{} and \ubase{}. The models explore different parametrizations to characterise the mode-mixed foregrounds and the 21-cm signal. Nevertheless, the noise is treated in the same manner for the five models. The component is estimated from the per-channel noise variance in Section~\ref{sec:noise model} and fitted by a white heteroscedastic noise kernel prior to nested sampling. The instrumental sensitivities are assumed accurately estimated. We then estimate the thermal noise by applying these sensitivities with a different random configuration in OSKAR to account for stochasticity. To capture the frequency dependence, the heteroscedastic kernel acts as an identity matrix with diagonal entries equal to the noise variance for the bandwidth, $\mathbf{K}_\mathrm{n}(\nu_i,\nu_j) = \sum_u \sigma^2_u(\nu)\delta_{ij}$. Additionally, each baseline bin $u$ is described by a noise kernel $\mathbf{K}_\mathrm{n}$ to characterise the potential noise variation across spatial scales. With the noise kernel calibrated to the estimated variances, the calibrated kernel $\mathbf{K}^\mathrm{cal}_\mathrm{n}$ is reinserted into the aggregate GPR model, $\mathbf{K} = \mathbf{K}_{21} + \mathbf{K}_\mathrm{fg} + \mathbf{K}^\mathrm{cal}_\mathrm{n}$. 

While modelling sky components, Model \anoise{}, \avarl{} and \mwedge\ incorporate multi-baseline effects of the mode-mixed foregrounds. The model parameters are also determined in each baseline bin to account for the variation of statistical properties of visibilities across different spatial scales. Model \manalytic{} describes the 21-cm signal with the exponential kernel, i.e. $\mathbf{K}_\mathrm{Mat\acute{e}rn}(\eta=\frac{1}{2})$ from Equation~\ref{eq:matern}, in place of the VAE kernel trained from Section~\ref{sec:kernel training}. \ubase{} retains the VAE kernel for the 21-cm signal, but models the foregrounds by treating the data across all the baselines with a single variance and length-scale. In \mwedge, the mode-mixed foreground length-scale is modelled to scale with baseline as,
\begin{eqnarray} \label{eq:wedge_parametrization}
    l_\mathrm{mix}(u) \propto \frac{1}{u\sin{\theta_\mathrm{mix}}+C_\mathrm{mix}},
\end{eqnarray}
where $u$ is the mean value of each baseline bin, $\bar{u}$ is the overall mean baseline length respectively, $\theta_\mathrm{mix}$ and $C_\mathrm{mix}$ are kernel parameters represent the wedge angle and wedge buffer respectively \citep{2025A&A...698A.186M}. Model \anoise{} extends \mwedge{} by describing the sky signals and noise as $\mathbf{K} = \mathbf{K}_{21} + \mathbf{K}_\mathrm{fg} + \alpha_\mathrm{n}\mathbf{K}_\mathrm{n}$, where $\alpha_\mathrm{n}$ is a noise-scaling parameter used to correct for any mismatch between the estimated and the actual noise variance. \avarl{} instead applies power-law scalings for both mode-mixed foreground variance and length-scale,
\begin{eqnarray}
    & \sigma^2_\mathrm{mix}(u) \propto (u/u_0)^{\alpha^\mathrm{var}_\mathrm{mix}}, \label{eq:power_law_mix_var} \\
    & l_\mathrm{mix}(u) \propto (u/u_0)^{\alpha^{l}_\mathrm{mix}}. \label{eq:power_law_mix_l}
\end{eqnarray}
Due to the power-law \rev{parametrization}, only the variance and length-scale in the first (reference) baseline bin, $\sigma^2_\mathrm{mix}(u_0)$ and $l_\mathrm{mix}(u_0)$, need to be estimated. Values at other baselines are therefore derived from the respective scaling relationships. These models are implemented using the ps\_eor python code. A summary of the five GPR models are listed in Table~\ref{tab:gpr_models} and their full parametrizations are given in Appendix~\ref{sec:gpr_model_parametrization_prior}.
\begin{table}
\centering
\caption{A summary of the candidate GPR models, listing the models, the baseline dependence of the mode-mixed foreground variance $\mathrm{\sigma^2_\mathrm{mix}}$ and length-scale $l_\mathrm{mix}$, the 21-cm kernel and noise scaling. The models, \mwedge{}, \anoise{} and \manalytic{}, describe length-scale using the ``wedge'' \rev{parametrization} (see Equation~\ref{eq:wedge_parametrization}). \avarl{} models both $\mathbf{\sigma^2_\mathrm{mix}}$ and $l_\mathrm{mix}$ with the power-law relationships described by Equation~\ref{eq:power_law_mix_var} and Equation~\ref{eq:power_law_mix_l} respectively. A dash (-) denotes the parameter has no baseline dependence. We also list the kernel functions used for modelling the 21-cm signal. ``VAE'' denotes the variational autoencoder kernel described in Section~\ref{sec:kernel training} and ``EXP'' denotes the exponential kernel function, ie a Mat\'ern kernel with $\eta = 1/2$. In \anoise{}, the estimated noise is scaled to fit the actual noise in the data as $\alpha_\mathrm{n}\mathrm{K_n}$ (see Section~\ref{sec:candidate_gpr_models} for details). Dashes here indicate no noise scaling applied.} \label{tab:gpr_models}
\renewcommand{\arraystretch}{1.5}
\setlength{\tabcolsep}{3pt}
\begin{tabular}{@{}ccccc@{}}
\toprule
\textbf{Model} & $\boldsymbol{\sigma^2}_\mathrm{mix}$ & $\boldsymbol{l}_\mathrm{mix}$ & $\mathbf{K}_{21}$ & \textbf{Noise scaling} \\
\midrule
\mwedge{} & $-$ & $\propto (u\sin{\theta_\mathrm{mix}}+C_\mathrm{mix})^{-1}$ & VAE & $-$ \\
\anoise{} & $-$ & $\propto (u\sin{\theta_\mathrm{mix}}+C_\mathrm{mix})^{-1}$ & VAE & $\alpha_\mathrm{n}\mathrm{K}_\mathrm{n}$ \\
\avarl{} & $\propto (u/u_0)^{\alpha^\mathrm{var}_\mathrm{mix}}$ & $\propto (u/u_0)^{\alpha^{l}_\mathrm{mix}}$ & VAE & $-$ \\
\manalytic{} & $-$ & $\propto (u\sin{\theta_\mathrm{mix}}+C_\mathrm{mix})^{-1}$ & EXP & $-$ \\
\ubase{} & $-$ & $-$ & VAE & $-$ \\
\bottomrule
\end{tabular}
\end{table}

% The noise can be estimated using various methods, such as time-differencing, frequency-differencing or using Stokes V flux that is circular and should contain the least information of the sky. In this case, we estimate the noise by using a different random seed from the injected thermal noise in the simulated sky data, by assuming a case where instrumental sensitivity is perfectly known.

\subsection{Bayesian framework} \label{sec:bayesian_framework}

The Bayesian evidence and posterior density distribution of kernel parameters for individual models are estimated by using a nested sampler PolyChord \citep{2015MNRAS.450L..61H,2015MNRAS.453.4384H} for the subsequent model selection and parameter estimation. The posterior distribution also provides \textit{maximum a posteriori} (MAP) estimates of the model parameters. This also enables a comparative analysis in 21-cm power reconstruction against the injected truth.

In Bayesian model comparison, global evidence $\mathcal{Z}$\footnote{Given that global evidence is often termed as marginal likelihood, we specifically refer to the global evidence as the likelihood function of the Gaussian Process Regression (GPR) model marginalised over the kernel parameters throughout the following discussion. Meanwhile, we use ``log-marginal likelihood (LML)'' to specifically denote the likelihood marginalised over all latent functions generated by Gaussian processes as described in Section~\ref{sec:gpr}.} is a statistical measure to quantify the probability density of observing the dataset $\mathbf{d}$, conditioned on the selected model $\mathcal{M}$,
\begin{eqnarray} \label{eq:global evidence}
    P(\mathbf{d}|\mathcal{M}) \equiv \mathcal{Z} = \int P(\mathbf{d}|\theta_\mathcal{M},\mathcal{M}) P(\theta_\mathcal{M}|\mathcal{M}) \mathrm{d}\theta_\mathcal{M}
\end{eqnarray}
where $\theta_\mathcal{M}$ represents the parameters of the model $\mathcal{M}$. The first term of the integral in Equation~\ref{eq:global evidence} is the likelihood function $\mathcal{L}(\theta) \equiv P(\mathbf{d}|\theta_\mathcal{M},\mathcal{M})$ that quantifies the probability of observing the data $\mathbf{d}$, under a set of parameters $\theta_\mathcal{M}$ that belong to a model $\mathcal{M}$. The second term is the prior density $\mathcal{\pi}(\theta) \equiv P(\theta_\mathcal{M}|\mathcal{M})$ that represents our initial knowledge about the model parameters before observations. The integral marginalises out the dependence of Bayesian evidence $\mathcal{Z}$ on the model parameters $\theta_\mathcal{M}$.

For each GPR model selected, we also estimate the posterior density distribution for the kernel parameters. According to Bayes' theorem, the posterior density is derived from the likelihood function $\mathcal{L}(\theta)$, prior density $\mathcal{\pi}(\theta)$ and model evidence $\mathcal{Z}$ by the expression 
\begin{eqnarray} \label{eq:posterior}
P(\theta_\mathcal{M}|\mathbf{d},\mathcal{M}) =
\frac{P(\mathbf{d}|\theta_\mathcal{M},\mathcal{M}) P(\theta_\mathcal{M}|\mathcal{M})}{P(\mathbf{d}|\mathcal{M})} \equiv \frac{\mathcal{L}(\theta)\mathcal{\pi}(\theta)}{Z}.
\end{eqnarray}
By applying the MAP estimates, the optimised GPR models are applied to the source-subtracted data from Section~\ref{sec:data processing} to further clean the residual foregrounds and extract the 21-cm signal. This will minimise the sky and instrumental contamination present in the radio observations and improve the reliability of the recovered cosmological signal. In addition, we also quantify the uncertainty for these parameters by analyzing their posterior distributions. The uncertainty is propagated through the subsequent analysis and allows us to assess its impact on the power spectrum measurements.

% Since we treat the noise as white noise, we can estimate the noise distribution from system temperature and hence sensitivity of the telescope. The noise is sampled from the same distribution but with a different random seed from the one used in Section~\ref{sec:noise model}. 

Among the five GPR models, ``\ubase{}'' does not account for spatial variations of sky signals across different baselines. The fewer degrees of freedom allows for a less flexible representation of the observed data. However, this also results in less complexity. We conduct comparisons between the five models using their Bayesian evidence $\mathcal{Z}$, root-mean-squared errors (\textit{RMSE}) and $z$-scores. The Bayesian evidence $\mathcal{Z}$ provides a global measure of how well a model explains the observed data while accounting for its complexity. In practice, we use Bayes factors, quantified as the ratio of the evidence between two competing models $\mathcal{M}_1$ and $\mathcal{M}_2$, to determine their relative merits, which is defined by,
\begin{eqnarray} \label{eq:bayes factor}
    B_{ij} = \frac{\mathcal{Z}_i}{\mathcal{Z}_j}.
\end{eqnarray}
In the case of large evidence values, we use logarithm of the Bayes factors in model comparisons, defined by,
\begin{eqnarray} \label{eq:log bayes factor}
    \mathrm{log}B_{ij} \equiv \Delta\mathrm{log}\mathcal{Z}_M = \mathrm{log}\mathcal{Z}_i - \mathrm{log}\mathcal{Z}_j.
\end{eqnarray}
Furthermore, for each GPR specification $M$ with kernels $\mathbf{K} = \mathbf{K}_\mathrm{21} + \mathbf{K}_\mathrm{fg} + \mathbf{K}_\mathrm{n}$, we construct a corresponding no-21 model $M_\mathrm{no-21}$ with $\mathbf{K} = \mathbf{K}_\mathrm{fg} + \mathbf{K}_\mathrm{n}$. These models without the 21-cm term therefore attempt to describe the sky signal with only foreground and noise. The Bayesian evidence for both models $M$ and their no-21 versions $M_\mathrm{no-21}$ are estimated using nested sampling, denoted by $\mathrm{log}\mathcal{Z}_M \equiv \mathrm{log}\mathcal{Z}(M)$ and $\mathrm{log}\mathcal{Z}_N \equiv \mathrm{log}\mathcal{Z}(M_\mathrm{no-21})$ respectively. By evaluating the log evidence difference between $M$ and $M_\mathrm{no-21}$, we can quantify the contribution of the 21-cm kernel $\Delta\mathrm{log}\mathcal{Z}_\mathrm{21}$, for each pair of models\footnote{It should be noted that the physical meaning of this log evidence difference $\Delta\mathrm{log}\mathcal{Z}_{21}$ differs from the meaning of the aforementioned log evidence difference $\Delta\mathrm{log}\mathcal{Z}_{M}$ between models $M$ even though they are both obtained by taking the difference of the Bayesian evidence. $\Delta\mathrm{log}\mathcal{Z}_{21}$ compares each model with and without the 21-cm kernel, while $\Delta\mathrm{log}\mathcal{Z}_{M}$ refers to the relative performance of the five primary models against a reference model.}. This difference also suggests if the observed sky can be described without 21-cm signal under a certain model.

In the following parameter estimation, we use \textit{RMSE} and $z$-scores to quantify the discrepancy between model predictions and observations. For our purposes, this is specifically defined as the difference between the extracted signal and the 21-cm signal observed from the simulation. Both quantities therefore measures the accuracy and reliability of signal recovery for a Bayesian GPR model, independent of model complexity. To assess the reliability of parameter estimation and therefore signal recovery, we also inspect the posterior distributions of the kernel parameters in the optimal model. A posterior distribution that deviates significantly from a normal distribution or exhibits abrupt cutoffs may introduce biases to estimation of the MAP parameters and their uncertainty. Additionally, we validate each model with Bayesian null tests to determine if accurate total-sky fits are achieved with biased sky components.

\section{Results} \label{sec:results}

\subsection{Bayesian evidence} \label{sec:bayesian_evidence}
\begin{figure*}
    \centering
    \includegraphics[width=\linewidth]{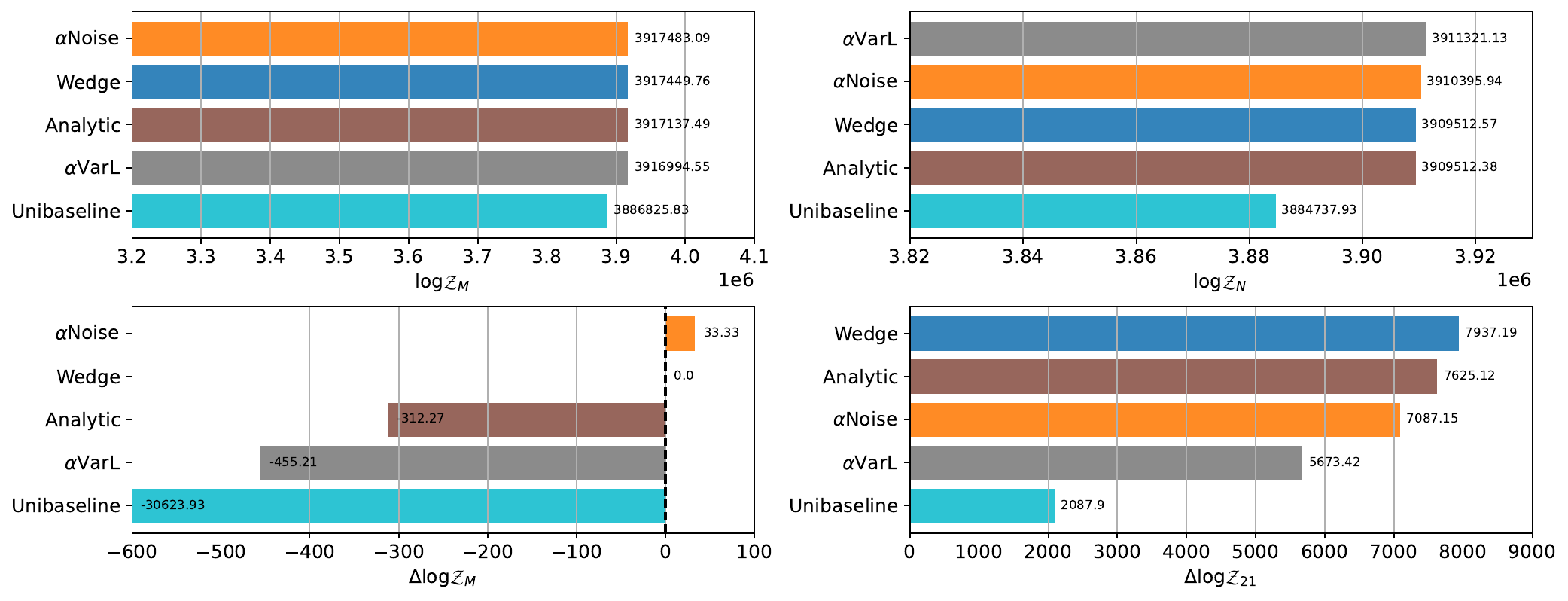}
    \caption{Bayesian statistical comparisons of the candidate GPR models. \textit{Upper left}: Log Bayesian evidence, $\mathrm{log}\mathcal{Z}_M$, fitted to the source-subtracted sky signal for \anoise{} (orange), \mwedge{} (blue), \manalytic{} (brown), \avarl{} (grey) and \ubase{} (teal). Larger $\mathrm{log}\mathcal{Z}_M$ indicates stronger overall support of the observed data after accounting for model complexity. \textit{Upper right}: Log evidence of the no-21 counterparts for the same dataset, $\mathrm{log}\mathcal{Z}_N$. The quantity measures how well each specification explains the observed signal without incorporating the cosmological component. \textit{Lower left}: Log Bayes factors relative to the reference model \mwedge, $\Delta \mathrm{log}\mathcal{Z}_M$, quantifying the relative support among the five primary models $M$. \rev{The axis is truncated to restrict the lower limit for readability.} \textit{Lower right}: Evidence gain from adding the 21-cm kernel, $\Delta\mathrm{log}\mathcal{Z}_\mathrm{21} = \mathrm{log}\mathcal{Z}_M - \mathrm{log}\mathcal{Z}_N$. \rev{A positive value indicates that a model including a 21-cm signal component is preferred relative to the equivalent model excluding that component.}} \label{fig:bayesian stats}
\end{figure*}

Based on the global log-evidence,  $\mathrm{log}\mathcal{Z}_M$, \anoise{} provides the best description for the observed data. The combined kernel of this model effectively reconstructs the total sky signal while maintaining a balanced model complexity. Since Model \anoise{} is built on Model \mwedge{} with an extra noise-scaling factor\rev{, this} parameter provides an additional degree of freedom that accounts for the difference between the estimated and actual noise, compared to \mwedge{}. Nevertheless, \anoise{} and \mwedge{} are also competing models, with the log Bayes factor $\mathrm{log}B_{ij} (\equiv \Delta\mathrm{log}\mathcal{Z}_M)$ between the two models, standing at \rev{$33.33$}, as shown in the \rev{lower} left panel of Figure~\ref{fig:bayesian stats}. This difference between the two models shows that incorporating more complicated noise amplitude variation in \anoise{} leads to an improvement in explaining the overall sky signal.

In addition, Figure~\ref{fig:bayesian stats} \rev{shows that} the Bayes factor between \avarl{} and \mwedge{} \rev{is} substantially more negative than that between \anoise{} and \mwedge{}. Model \avarl{} is less supported by \rev{an evidence difference $\Delta\mathrm{log}\mathcal{Z}_M$ of 455.21} compared with the reference model \mwedge{}. This indicates modelling the spatial variation of sky signals with power-law relationships is less supported by the observed data than modelling with wedge \rev{parametrization}. \rev{The $\mathrm{log}\mathcal{Z}_M$ of \ubase{} is less supported by 30623.93 than the reference.} By comparing \avarl{} and \ubase{}, however, these spectral parameters in \avarl{} provide additional degrees of freedom and allow for a more flexible description of the dataset despite introducing complexity to the model. \rev{In terms of the significance of the 21-cm component, the evidence gains $\Delta\mathrm{log}\mathcal{Z}_{21}$ are positive for all five models, as shown in the lower right panel of Figure~\ref{fig:bayesian stats}. The observed sky data is less favoured while modelling with $\mathbf{K} = \mathbf{K}_\mathrm{fg} + \mathbf{K}_\mathrm{n}$.} The contribution from the 21-cm component is the most significant in \mwedge{}, with an evidence gain \rev{$\Delta\mathrm{log}\mathcal{Z}_{21} = 7.94\times10^3$}.

Among the five models, the posterior distributions of the kernel parameters are all found unimodal and with well-defined credible intervals, \rev{as presented in Appendix~\ref{sec:gpr_model_parametrization_posterior}}. In particular, the amplitudes of the two foreground components in the models, the intrinsic 
$\sigma^2_\mathrm{int}$ and the mix term $\sigma^2_\mathrm{mix}$ exhibit weak correlations in their joint posterior contour. This indicates that the data and the chosen kernel \rev{parametrization} cleanly separate the two contributions: the intrinsic component is primarily constrained by the spectrally smooth covariance of the intrinsic foregrounds, while $\sigma^2_\mathrm{mix}$ is governed by the more complicated, wedge-like and baseline-scaling covariance introduced by chromatic mode-mixing. The lack of multimodality also suggests stable evidence estimates and weak parameter correlations that may lead to degeneracies. This provides an extra layer of reliability to the subsequent model comparison and the reconstruction of the 21-cm signal obtained from the MAP kernel parameters. \rev{We note that the noise scaling factor $\alpha_\mathrm{n}$ in \anoise{} stands at $1.014^{+0.002}_{-0.002}$, instead of exactly unity, in the absence of frequency-uncorrelated noise in our simulations. Although source subtraction can inadvertently remove background noise, incorrectly predict source flux and introduce frequency-uncorrelated artefacts to the sky signals and some of the 21-cm signal and foregrounds may be treated as noise in the GPR analysis, the observed 
mismatch by $\sim1\%$ suggests these effects are insignificant. We find this deviation from unity falls within the limit of statistical sampling error. We test this by fitting multiple random realisations of noise-only synthetic data from the same Gaussian distribution with $\alpha_\mathrm{n}\mathbf{K}^\mathrm{cal}_\mathrm{n}$ and estimate the scaling factor $\alpha_\mathrm{n}$ from the injected noise in our simulation. We observed a spread of values around 1.0, resulting in an average absolute deviation of approximately 2\%. Therefore, we conclude that the 1.4\% deviation from the standard deviation of the normal distribution of the noise realisations in our primary analysis is not improbable given this distribution.}

In addition to the models discussed above, we also assess the no-21 models $M_\mathrm{no-21}$ as to whether the data can be explained without including the 21-cm signal component. These models are obtained by retaining only the kernel functions describing foregrounds and instrumental noise while removing the dedicated kernel for the 21-cm signal. We find that all the models demonstrate positive $\Delta\mathrm{log}\mathcal{Z}_\mathrm{21}$ that indicate a statistical preference for including a dedicated 21-cm kernel component. The $\mathrm{log}\mathcal{Z}_N$ of \mwedge{} and \manalytic{} remain the same because both models have the same foreground kernel components. The only difference lies in their treatment of the 21-cm signal: while \mwedge{} models the component using the VAE kernel, \manalytic{} uses the Matérn exponential kernel. Among all models, \mwedge{} shows the strongest support for the presence of the 21-cm signal component, \rev{followed by \manalytic{}, \anoise{}, \avarl{} and \ubase{} that are less supported by 3.9\%, 10.7\%, 28.5\% and 73.7\% than Model \mwedge{}, respectively, in $\Delta \mathrm{log}\mathcal{Z}_{21}$.}

% $7.15 \times 10^3$, $6.51 \times 10^3$, $5.74 \times 10^3$ and $2.14 \times 10^3$ respectively.

\subsection{21-cm signal reconstruction}
\begin{figure*}
    \centering
    \includegraphics[width=\linewidth]{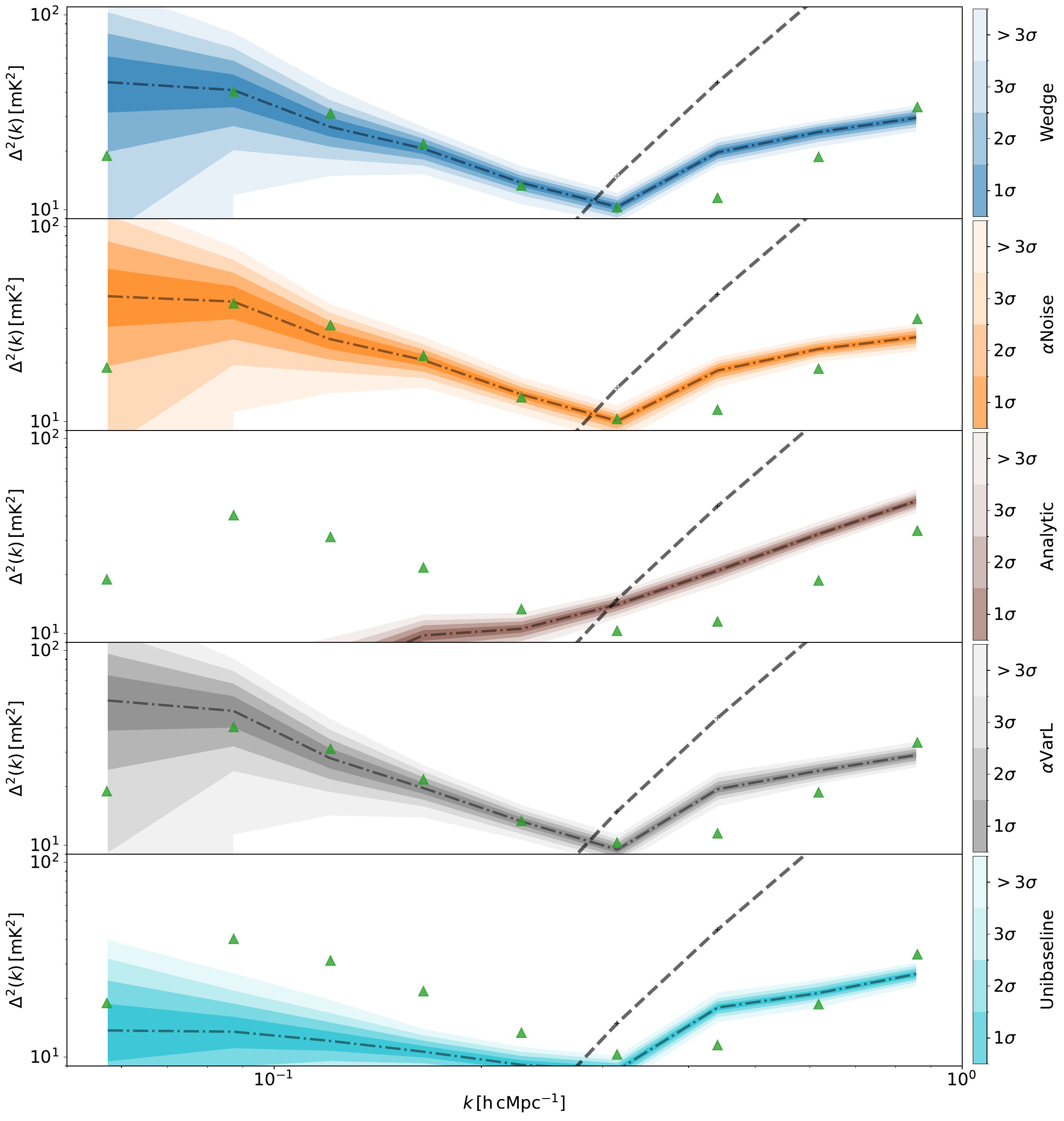}
    \caption{Recovered 21-cm power spectra by the GPR models. From top to bottom, the five panels show the spherical dimensionless 1D power spectra of the 21-cm signal, $\Delta^2_\mathrm{rec}(k)$, reconstructed by the GPR models \mwedge{} (blue), \anoise{}{} (orange), \manalytic{} (brown), \avarl{} (grey) and \ubase{} (teal), compared with the injected (``true'') 21-cm signal (green triangles) $\Delta^2_\mathrm{21}(k)$ and thermal noise level $\Delta^2_\mathrm{n}(k)$ (dashed line). For each model, the dash-dot curve shows the posterior-predictive median, and the shaded bands denote the $1\sigma$, $2\sigma$, $3\sigma$ and $>3\sigma$ credible intervals. The predictive distribution of the recovered $\Delta^2(k)$ for each model is estimated by using the procedures in Section~\ref{sec:z_scores}.} \label{fig:gpr_models_ps1d}
\end{figure*}

Figure~\ref{fig:gpr_models_ps1d} shows the recovered power spectra of the 21-cm signal from the sky signal for the five GPR models, in reference with the injected true 21-cm power. In general, \mwedge{}, \anoise{} and \avarl{} exhibit comparable accuracy in reconstructing the 21-cm power spectra, delivering the lowest residuals across the intermediate modes for $0.087 < k < 0.32\ \mathrm{h\ cMpc}^{-1}$. Within this range, the recovered power mostly passes through the injected 21-cm power within the $2\sigma$ credible envelope. The residuals between the recovered power and the injected 21-cm signal are shown in the upper panel of Figure~\ref{fig:gpr_models_res_z_scores}. \rev{We find that the mismatch becomes more pronounced at large $k$ due to the increased noise level in this regime. Because the VAE kernel is trained on 21-cm power spectra from 21cmFAST, it enforces a functional prior on the shape of the predicted signal, and measurements across different $k$-modes are not independent. While the kernel accurately fits the high-signal-to-noise data at small $k$, this global shape constraint forces the predicted power at large $k$ to compromise, lying between the true 21-cm signal and the dominant noise.} \mwedge{} attains the best local recovery at $k = 0.32\ \mathrm{h\ cMpc}^{-1}$, predicting a 21-cm power of \rev{$10.33\ \mathrm{mK}^2$}. The local recovery shows a slight overestimation \rev{by} a fractional difference of \rev{$0.10\%$}, compared to the injected truth of $10.32\ \mathrm{mK}^2$. Similar to \mwedge{}, the closest match for \anoise{} occurs at the same $k$-mode with a fractional difference of \rev{$-0.24\%$}, suggesting a conservative estimate of the injected 21-cm power. Model \avarl{}, on the other hand, performs the best at $k = 0.23\ \mathrm{h\ cMpc}^{-1}$, slightly underestimating the 21-cm power by \rev{$0.092\ \mathrm{mK}^2$} and reaching a fractional difference of \rev{$-0.69\%$}. More significant differences are noticed for \ubase{} and \manalytic{}. Both models systematically underestimate the cosmological power on large scales for $k < 0.23\ \mathrm{h\ cMpc}^{-1}$ as shown in Figure~\ref{fig:gpr_models_ps1d}. This difference remains pronounced even at their optimal recovery $k$-modes, \ubase{} by \rev{$-16.7\%$} at $k = 0.32\ \mathrm{h\ cMpc}^{-1}$ and \manalytic{} \rev{by $-20.6\%$} at $k = 0.23\ \mathrm{h\ cMpc}^{-1}$, respectively. These comparisons therefore highlight the local reconstruction capability of \anoise{} in terms of fractional difference from the injected 21-cm signal, with progressively lower accuracy demonstrated by \mwedge{}, \avarl{}, \ubase{}, and \manalytic{}.

% In terms of signal recovery, Model \mwedge{} evidently outperforms other test models in the accuracy of separating the 21-cm signal from residual astrophysical foregrounds and thermal noise, by achieving the lowest \textit{RMSE} at $8.27 \times 10^6$ (see Figure~\ref{fig:gpr_models_cross_metrics}). Models \anoise{}, \avarl{}, and \ubase{} follow \mwedge{} in performance, exhibiting \textit{RMSE} values greater than \mwedge{} by fractional differences of 12.0\%, 35.4\%, and 36.6\%, respectively. Among the evaluated models, Model \manalytic{} yields the highest \textit{RMSE} that exceeds that of \mwedge{} by a factor of 2.24. The difference in \textit{RMSE} indicates a less accurate extraction of the 21-cm signal by using analytical kernels.

% \mwedge{} tends to edge out the others by explicitly capturing wedge-like, multi-baseline covariance, and \anoise{} matches this performance while using a single noise-scaling parameter to absorb small variance mismatches without leaking into the astrophysical components. \avarl{} is competitive but shows a mild low-bias at the largest scales, consistent with the amplitude–coherence trade-off between its power-law indices for variance and length-scale.

\subsection{$z$-scores} \label{sec:z_scores}
To assess the reliability of the recovered 21-cm signals for different GPR models, we evaluate $z$-scores that explicitly account for predictive accuracy and uncertainty. The $z$-score at each sampled Fourier mode $k_i$ is be expressed by,
\begin{eqnarray} \label{eq:z_i}
z_i =
\begin{cases}
\dfrac{\Delta^{2}_{i,\mathrm{rec}} - \Delta^{2}_{i,21}}{\sigma_i^{-}}, & \text{if }\ \Delta^{2}_{i,\mathrm{rec}} \ge \Delta^{2}_{i,21},\\[6pt]
\dfrac{\Delta^{2}_{i,\mathrm{rec}} - \Delta^{2}_{i,21}}{\sigma_i^{+}}, & \text{if }\ \Delta^{2}_{i,\mathrm{rec}} < \Delta^{2}_{i,21},
\end{cases}
\end{eqnarray}
where $\Delta^{2}_\mathrm{i,rec}$ is \rev{the median of the possible dimensionless 1D power spectra} recovered by a GPR model, $\Delta^{2}_\mathrm{i,21}$ is the \rev{median of the} injected 21-cm power, $\sigma_i^{+}$, and $\sigma_i^{-}$ denotes the upper and lower $1\sigma$ predictive uncertainty \rev{on the power} at sampled $k$-mode $k_i$. We summarise the performance of each model by using the \textit{RMS} $z$-scores averaged over the $N$ $k$-modes sampled in the spherical power-spectrum estimation:
\begin{eqnarray} \label{eq:averaged_z_score}
    z = \sqrt{\frac{\sum_{i}^N{z_i}^2}{N}}.
\end{eqnarray}

\begin{figure*}
    \centering
    \includegraphics[width=\linewidth]{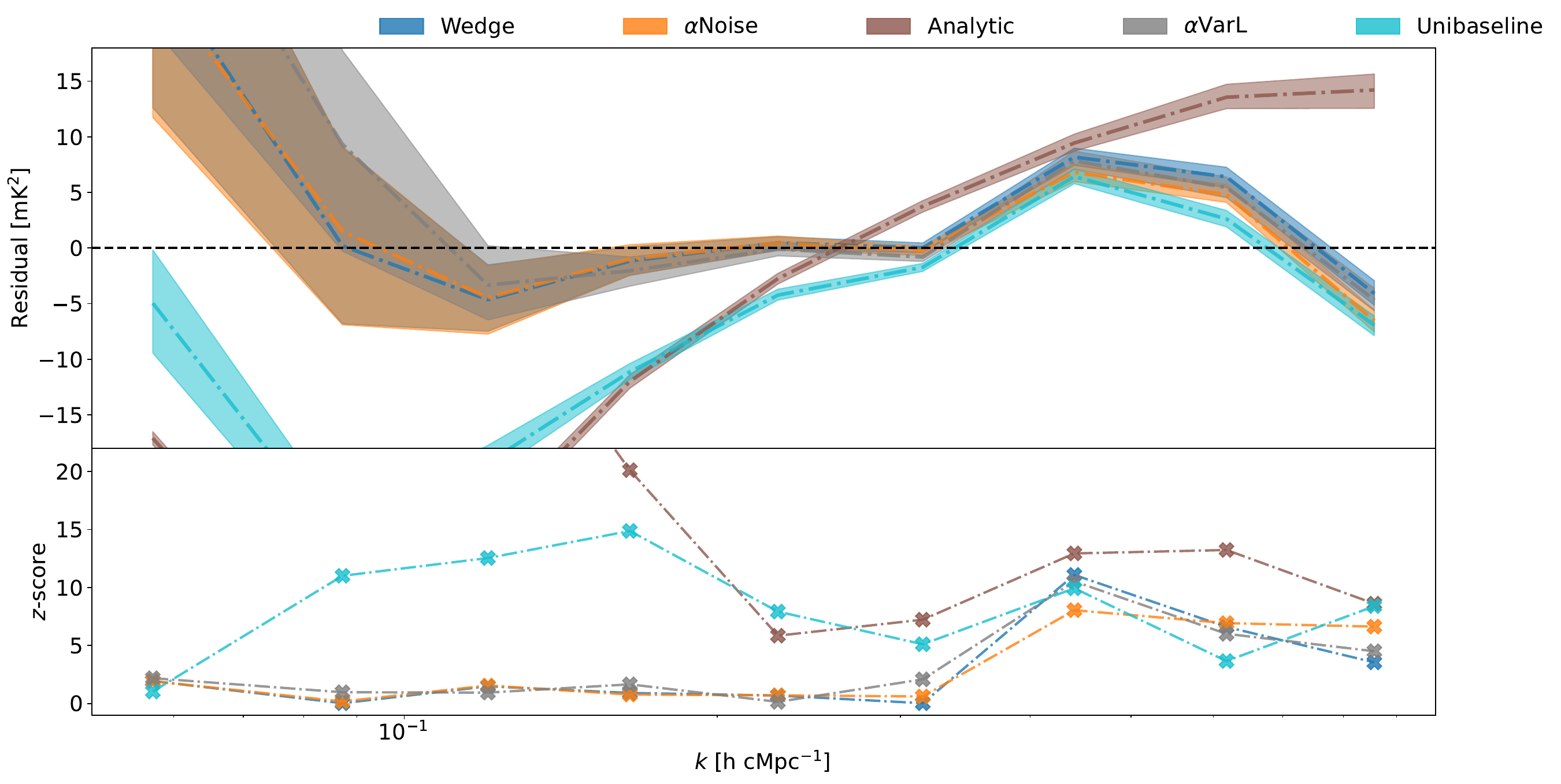}
    \caption{Power residuals and $z$-scores over the sampled $k$-modes. \textit{Upper panel}: The residuals $\Delta^2_\mathrm{rec}(k) - \Delta^2_\mathrm{21}(k)$ as a function of the 1D Fourier modes $k$ for the five GPR models by using the same colours as in Figure~\ref{fig:gpr_models_ps1d}. The dashed horizontal line marks the zero residual, with the curves above(below) indicates an overestimation(underestimation) of the cosmological signal. The shaded area indicates the $1\sigma$ credible interval. \textit{Lower panel}: The per-mode $z_i(k)$ values that quantify the normalised residual at each $k$-mode by the corresponding asymmetric $1\sigma$ uncertainty (see Equation~\ref{eq:z_i}). Due to the sign change of asymmetric uncertainty, $z_i$ can only be positive and lower $z_i$ indicates more reliable local recovery of the 21-cm component at Fourier mode $k_i$.} \label{fig:gpr_models_res_z_scores}
\end{figure*}

To estimate the uncertainty for each model $\sigma_i^{\pm}$, we follow and extend the procedures introduced in \citet{2025A&A...698A.186M}. First, we draw 50 samples from the joint posterior of kernel hyperparameters with respect to their nested sampling weights. The predictive GP
mean $\langle \mathbf{f_{21}} \rangle$ and covariance $\mathrm{cov}(\mathbf{f_{21}})$ are computed for each parameter sample by using Equation~\ref{eq:gp_mean_cov}. Given a specific set of parameters, we randomly sample 50 realisations for the 21-cm signal from $\mathbf{f}_{21} \sim \mathcal{N}(\langle \mathbf{f}_{21} \rangle,\mathrm{cov}(\mathbf{f}_{21}))$ to capture the intrinsic GP stochasticity. For each realisation of the 21-cm signal, we estimate the binned spherical 1D power spectrum $\Delta^2(k)$ using Equation~\ref{eq:dimensionless_ps1d} and its per-bin uncertainty $\sigma_\mathrm{s}(k)$ from the sample variance of values entering each $k$ bin. To propagate the power spectrum estimation error,  we draw 100 samples per bin from $\mathcal{N}(\langle \Delta^2(k) \rangle,\sigma_\mathrm{s}(k))$, yielding an ensemble of power spectra for that realisation. Aggregating over all the realisations from posterior, GP and power spectrum binning produces a predictive distribution with $50 \times 50 \times 100$ samples of $\Delta^2(k)$ at each $k$ bin. The 16 and 84 percentiles of this distribution are therefore calculated as the empirical $1\sigma$ credible interval for a given GPR model. We find these sample sizes yield stable posterior estimates of $\Delta^2(k)$ while incurring manageable computation runtime. \rev{Here the uncertainty $\sigma_i^{\pm}$ in Equation~\ref{eq:z_i} quantifies errors in the model prediction on the ground truth of the 21-cm signal. This term incorporates both the uncertainty propagated from the posterior samples of model parameters as shown in Appendix~\ref{sec:gpr_model_parametrization_posterior} and the sampling variance of the 21-cm power spectrum for the following model comparison. To perform the subsequent model comparison, this uncertainty only concerns model ignorance and does not include thermal noise sampling variance.} The lower panel of Figure~\ref{fig:gpr_models_res_z_scores} shows the $z$-scores for the five models at sampled Fourier modes. 

From the lower panel of \rev{Figure~\ref{fig:gpr_models_res_z_scores}}, \anoise{}, \mwedge{}, and \avarl{} exhibit similar $z$-score profiles over the entire $k$-range. \rev{The three models achieve the lowest $z$-scores at $k = 0.32\ \mathrm{h\ cMpc}^{-1}$, $k = 0.087\ \mathrm{h\ cMpc}^{-1}$ and $k = 0.23\ \mathrm{h\ cMpc}^{-1}$ respectively, with a noticeable increase towards higher $k$ until $k > 0.32\ \mathrm{h\ cMpc}^{-1}$}. These models are largely consistent with each other at small and medium scales for \rev{$k < 0.23\ \mathrm{h\ cMpc}^{-1}$}, while the differences become more pronounced at larger scales. While only considering the $k$-modes where the 21-cm power is \rev{above} the noise floor, the \textit{RMS} z-scores of \anoise{}, \mwedge{} and \avarl{} stand at \rev{$1.20$, $1.20$ and $1.37$} respectively, for $k < 0.23\ \mathrm{h\ cMpc}^{-1}$. In addition, \manalytic{} have the largest $z$-scores over nearly the entire $k$-range, except for a significant drop due to its improved predictive residuals \rev{until $k > 0.23\ \mathrm{h\ cMpc}^{-1}$}. \ubase{} also shows large $z$-scores in general, however its performance begins to approach that of the best three models from $k = 0.44\ \mathrm{h\ cMpc}^{-1}$ onward.

% indicating more reliable signal reconstruction at large scales.

\subsection{Cross-metric comparison} \label{sec:cross_metric_comparison}

\begin{figure*}
    \centering
    \includegraphics[width=\linewidth]{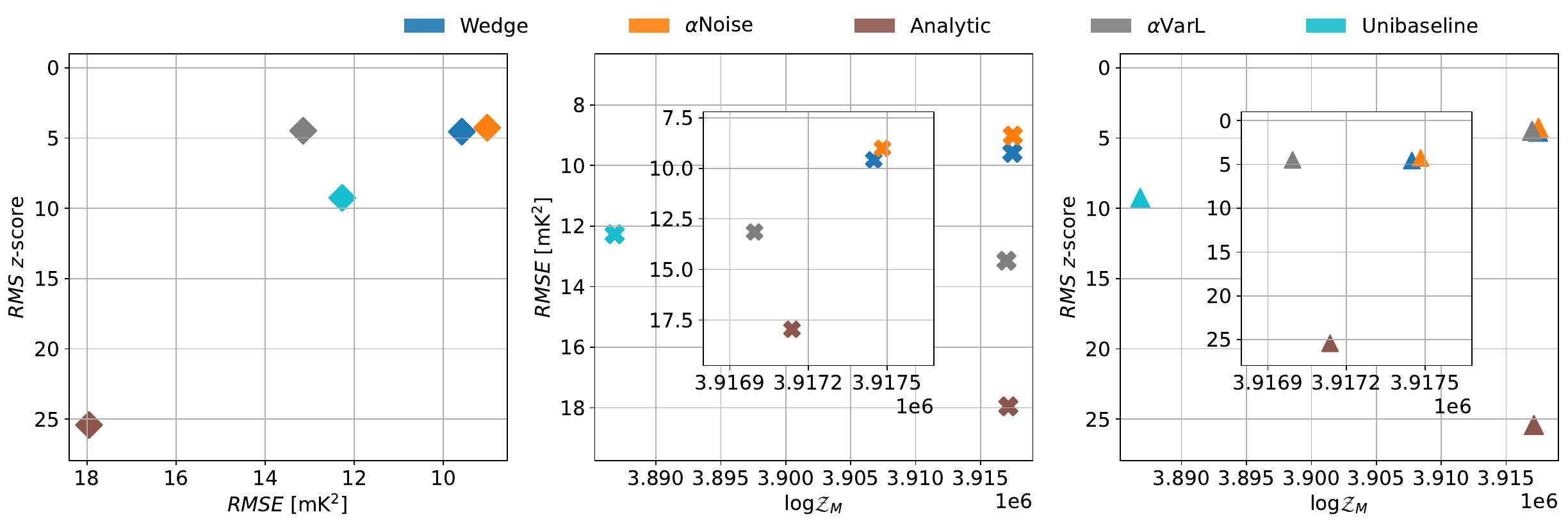}
    \caption{Cross-metric comparisons of the GPR model performance between $k$-mode averaged \textit{RMSE}, \textit{RMS} $z$-scores and log model evidence $\mathrm{log}\mathcal{Z}_M$. \textit{Left panel} shows the \textit{RMS} $z$-scores of the GPR models against \textit{RMSE}. As both metrics are derived from residuals between the recovered and injected 21-cm power, the models generally follow a positive trend. \textit{Middle panel} shows the relationship between \textit{RMSE} and $\mathrm{log}\mathcal{Z}_M$, comparing the 21-cm reconstruction accuracy and aggregate predictive adequacy. \textit{Right panel} shows the \textit{RMS} $z$-scores against $\mathrm{log}\mathcal{Z}_M$. The comparison suggests the relationship between the relative data support and recovery reliability of the 21-cm component. The vertical axes are inverted for the \textit{RMSE} and \textit{RMS} $z$-scores to place better models closer to the top right corner of the figures. Meanwhile, insets in the middle and right panels exclude \ubase{} due to the notably small $\mathrm{log}\mathcal{Z}_M$, to improve visual separation of the remaining models.} \label{fig:gpr_models_cross_metrics}
\end{figure*}

Figure~\ref{fig:gpr_models_cross_metrics} summarises and compares the three metrics used in the analysis: the log Bayesian evidence $\mathrm{log}\mathcal{Z}_M$, the \textit{RMSE} averaged over all the sampled $k$-mode, and the $k$-mode averaged \textit{RMS} $z$-score. Naturally, \textit{RMSE} and \textit{RMS} $z$-score are generally positively correlated since both quantities are functions of the residuals between the recovered and injected 21-cm power. As an overview, \anoise{} achieves the lowest \textit{RMS} $z$-score of \rev{$4.27$ and the lowest \textit{RMSE} of $9.02\ \mathrm{mK^2}$}. A notable exception is seen for \avarl{} that attains lower \textit{RMS} $z$-score than \ubase{} despite having larger \textit{RMSE}. This implies that \avarl{} has larger predictive uncertainty and thus yields smaller $\sigma$-normalised difference while in fact having a larger residual. The log evidence $\mathrm{log}\mathcal{Z}_M$ does not show a monotonic relationship with \textit{RMSE} or $z$-score. While $\mathrm{log}\mathcal{Z}_M$ measures the overall model predictivity of the data, we use \textit{RMSE} and $z$-scores to describe the recovered 21-cm component specifically. This therefore suggests a sign for one or more components that may be biased even if they provide accurate fit to the total sky intensity. We will have a detailed discussion on this inconsistency between the metrics and provide a validation test of the biased recovery in the following section.

The left panel of Figure~\ref{fig:gpr_models_cross_metrics} shows that Model Analytic has the highest power spectrum residuals, with an \textit{RMSE} of \rev{$18.0\ \mathrm{mK^2}$}, and most statistically significant errors, with an \textit{RMS} z-score averaged over the sampled k-modes of \rev{$25.4$}. The middle panel of Figure~\ref{fig:gpr_models_cross_metrics} shows that \ubase{} attains the lowest $\mathrm{log}\mathcal{Z}_M$, yet attains an averaged \textit{RMSE} of \rev{$12.3\ \mathrm{mK^2}$} that suggests better reconstruction of the 21-cm signal than \avarl{} and \manalytic{}. A similar pattern appears in the right panel: \ubase{} reaches an \textit{RMS} $z$-score of \rev{$9.26$}, implying higher reconstruction accuracy and reliability than \manalytic{}. However, \manalytic{} yields a $\mathrm{log}\mathcal{Z}_M$ being \rev{$0.78\%$} higher than \ubase{}, just exceeding \anoise{} and \mwedge{}. Moreover, while \avarl{} predicts a cosmological signal that departs further from the \rev{injected signal}, the $z$-score profiles of \mwedge{}, \anoise{} and \avarl{} are broadly similar due to the larger predictive uncertainty of \avarl{}.

% \begin{figure*}
%     \centering
%     \includegraphics[width=\linewidth]{Figs/gpr_models_null_rec.pdf}
%     \caption{The $1\sigma$ upper limits of the spherical 1D power spectra of the 21-cm signal $\Delta^2_{21}(k)$ predicted by the GPR models $M$, fitted to the sky data with the cosmological signal removed. The predicted power uses the same colour scheme as Figure~\ref{fig:gpr_models_ps1d}. The green line and triangles mark the ``true'' power of the 21-cm signal for reference. The dahsed line indicates the thermal noise level. Across the entire $k$-range, all the models, except \avarl{}, produce upper limits that lie below the reference 21-cm signal. Nevertheless, \avarl{} yields the weakest power $\Delta^2_{21}(k) \simeq 0.10\ \mathrm{mK^2}$ in the noise-dominated regime where thermal noise dominate for $k > 0.30\ \mathrm{h\ cMpc}^{-1}$.} \label{fig:gpr_models_null_rec}
% \end{figure*}

\subsection{Bayesian null test} \label{sec:bayesian_null_test}

Bayesian evidence measures the predictive adequacy given the aggregate model ($\mathbf{K} = \mathbf{K}_\mathrm{21} + \mathbf{K}_\mathrm{fg} + \mathbf{K}_\mathrm{n}$) and marginalizing over kernel parameters. \textit{RMSE} and $z$-scores, on the other hand, quantify the fidelity of the 21-cm component specifically. In multi-component settings, a model can obtain high evidence by fitting the total sky well, even if the decomposition is biased. This may ultimately lead to low reconstruction reliability for the predictive component of primary interest (i.e. the 21-cm signal). Model comparisons using Bayes factors can only rank models by overall predictive adequacy; however, it cannot by itself guarantee unbiased component recovery. As discussed in Section~\ref{sec:cross_metric_comparison} and shown in Figure~\ref{fig:gpr_models_cross_metrics}, comparisons, such as that between \avarl{} and \mwedge{}, are straightforward - the aggregate model with higher evidence is made of components that more accurately describe the signal of primary interest (i.e. the 21-cm signal). However, pairwise comparisons like \ubase{} versus \avarl{}(or \manalytic{}), fall into a more complicated scenario. Model \avarl{}(\manalytic{}) provides a better description of the overall data, yet deliver less accurate reconstruction to the component of interest. This indicates a risk of biased component recovery, where foreground and noise structures may be partially absorbed by the 21-cm term. This motivates the following test to validate if an accurate fit is achieved for the total sky intensity by biasing the inferred 21-cm amplitude. For this purpose, we use a Bayesian Null Test Evidence Ratio-based validation framework (BaNTER), introduced in \citet{2025MNRAS.541.2262S} to identify and exclude problematic models in this context. While \citet{2025MNRAS.544.2340S} demonstrates its application in global 21-cm cosmology, here we apply the BaNTER validation to interferometric observations for the first time.

% Model comparisons utilizing Bayes factors however cannot identify such problematic models.

% As shown in Figure~\ref{fig:gpr_models_null_rec}, applying the models $M$ to $\mathbf{d}_\mathrm{no-21}$ yields upper constraints on the 21-cm power that would arise if any residual foreground and noise structures were erroneously absorbed by the 21-cm kernel. Results from Figure~\ref{fig:gpr_models_null_rec} suggest that \avarl{} shows a sign of overestimating the cosmological signal on the largest scales for $k < 0.09\ \mathrm{h\ cMpc}^{-1}$. The $1\sigma$ upper limits predicted by the model are above the reference 21-cm signal by $257\ \mathrm{mK}^2$ and $117\ \mathrm{mK}^2$ at $k =0.057\ \mathrm{h\ cMpc}^{-1}$ and $k =0.087\ \mathrm{h\ cMpc}^{-1}$ respectively. On the other hand, \mwedge{} and \anoise{} delivers much weaker power, approximately one order of magnitude lower than the reference over the $k$-range. Nevertheless, \manalytic{} shows a more ideal case by predicting power below the noise floor, and does not prompt spurious detections at all scales. 

\begin{figure}
    \centering
    \includegraphics[width=\linewidth]{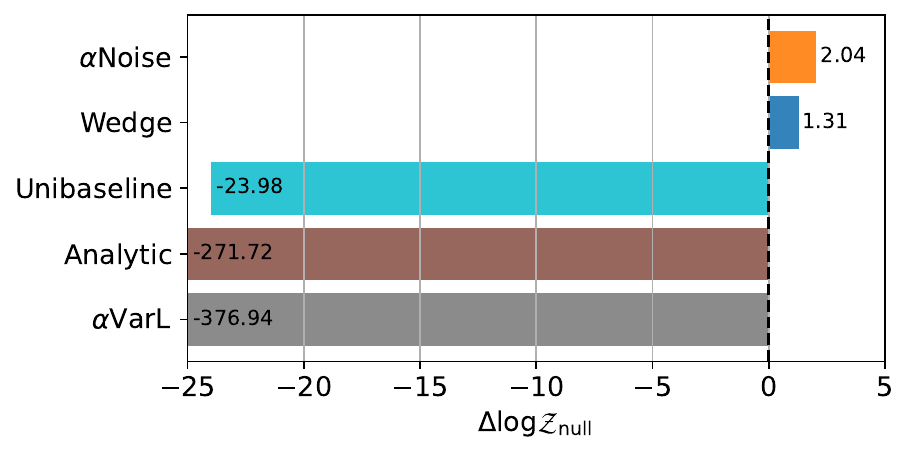}
    \caption{The log evidence difference $\Delta\mathrm{log}\mathcal{Z}_\mathrm{null} = \mathrm{log}\mathcal{Z}'_N - \mathrm{log}\mathcal{Z}'_M$, for the five GPR configurations. The difference compares the evidence for a model without a 21-cm component ($M_\mathrm{no-21}$) against the full model ($M$), evaluated on the 21-cm absent sky data $\mathbf{d}_\mathrm{no-21}$. The vertical dashed line marks $\Delta\mathrm{log}\mathcal{Z}_\mathrm{null} = 0$. Positive values indicate that the data correctly favours $M_\mathrm{no-21}$ over $M$. The mild positive offsets for \mwedge{} and \anoise{}, at 1.37 and 0.82 respectively, suggest they remain unbiased in this test, while the other configurations are shown to be strongly biased towards signal-detecting models when no cosmological signal is present. The lower end of $\Delta\mathrm{log}\mathcal{Z}_\mathrm{null}$ is capped at $-25$ for readability.}
    \label{fig:gpr_models_null_logZ}
\end{figure}

The validation test is conducted with a validation dataset, $\mathbf{d}_\mathrm{no-21} = \mathbf{f}_{\mathrm{fg}} + \mathbf{n}$. The idea is to validate each model on the data that contains only the nuisance signals ($\mathbf{f}_\mathrm{fg}$) and noise ($\mathbf{n}$), with the signal of interest absent. We estimate the Bayesian evidence for the full models $M$ and the no-21 models $M_\mathrm{no-21}$ on the 21-absent data $\mathbf{d}_\mathrm{no-21}$. The evidence of $M$ and $M_\mathrm{no-21}$ estimated on $\mathbf{d}_\mathrm{no-21}$ is denoted as $\mathcal{Z}'_M$ and $\mathcal{Z}'_N$ respectively. To prevent the zero-mean GP assumption from biasing the evidence, we mean-centre the data along the frequency axis before the evidence estimation. By using Equation~\ref{eq:log bayes factor}, the Bayesian null log evidence difference is expressed as $\Delta\mathrm{log}\mathcal{Z}_\mathrm{null} = \mathrm{log}\mathcal{Z}'_N - \mathrm{log}\mathcal{Z}'_M$. Under a null condition where no 21-cm signal is present in the observed data, a validated model does not gain evidence by adding the 21-cm term $\mathbf{K}_{21}$, i.e. $\Delta\mathrm{log}\mathcal{Z}_\mathrm{null} > 0$. From Figure~\ref{fig:gpr_models_null_logZ}, the evidence-based null test splits the models into two categories. \rev{\anoise{} and \mwedge{} show slight evidence gains by $2.04$ and $1.31$ while excluding the 21-cm component from the models.} The models that only contain nuisance components $M_\mathrm{no-21}$ are correctly favoured by the validation dataset $\mathbf{d}_\mathrm{no-21}$ for both configurations. By contrast, \ubase{}, \manalytic{} and \avarl{} fail the test with their increasingly negative $\Delta\mathrm{log}\mathcal{Z}_\mathrm{null}$. This signifies that including $\mathbf{K}_{21}$ results in a spurious increase in model evidence rather than the expected penalty. The 21-cm components in these models are therefore more likely to learn the residual errors or unmodelled structures from the nuisance components (i.e. $\mathbf{K}_\mathrm{int}$, $\mathbf{K}_\mathrm{mix}$ and $\mathbf{K}_\mathrm{n}$), due to imperfect \rev{parametrization}.

% In this work, GPR models are built based on Equation~\ref{eq:eor_jointdis}, that assumes a zero mean function. This joint
% probability density distribution is a common practice in foreground mitigation and EoR signal recovery \citep{2018MNRAS.478.3640M,2020MNRAS.493.1662M,2024MNRAS.527.3517M}. The GP will only explain the trend using the kernel functions. While this can define each component entirely by the kernels, this may lead to inefficient description of the dataset, which will be captured in Bayesian evidence. While a detailed discussion of mean functions used in the GPR is beyond the scope of this work, we work this around this by preprocessing the visibility data by removing the mean along the frequency axis before estimating the model evidence. 

% \begin{table}
% \centering
% \caption{} \label{tab:log_z_diff_no_21}
% \begin{tabular}{cc}
% \hline
% \hline
% Model & $\Delta\mathrm{log}\mathcal{Z}_\mathrm{null}$ \\
% \hline
% \mwedge{} & -1.37 \\
% \anoise{} & -0.082 \\
% \ubase{} & 24.3 \\
% \manalytic{} & 273.7 \\
% \avarl{} & 380.3 \\
% \hline
% \hline
% \end{tabular}
% \end{table}

\section{Conclusion} \label{sec:conclusion}
We present a comprehensive Bayesian framework to evaluate the performance of different GPR foreground mitigation models for the simulated SKA-Low observations. The observations are simulated with OSKAR by incorporating realistic sky models and beam responses that span a portion of SKA-Low bandwidth from 122 to 134 MHz. The sky models include galactic and extragalactic point sources from the GLEAM survey, diffuse emission from the GSM package, the 21-cm signal generated with 21cmFAST, and thermal noise equivalent to 1000h observation. The radio sky is continuously tracked for 4 hours on a single night under ideal weather conditions. A total of 1440 observations are conducted, each integrated over 10 seconds. The OSKAR simulations utilise a single station beam pattern and duplicated it across all 512 SKA-Low stations. 

With both natural and uniform weighting schemes, we produce interferometric images and apply the CLEAN algorithm to reduce dirty beam side-lobes using WSCLEAN. We find that uniform-weighted, deconvolved images excel at detecting and characterizing point radio sources, benefiting from their suppressed short baselines. \rev{Meanwhile, compared to uniform weighting, natural weighting preserves more diffuse structures due to the maximised sensitivity \footnote{GPR is applied to the PSF-corrected dataset and should theoretically yield equivalent results for both natural and uniform weighting according to Equation~\ref{eq:flux density} and Equation~\ref{eq:psf}. In practice, however, computational constraints limit PSF images to a specific FoV rather than the full sky, meaning they cannot capture all information from the ungridded uv data. The true PSF is therefore not fully recoverable from Equation~\ref{eq:psf}. This information loss occurs in both weighting schemes, however, uniform weighting features an additional weighting function accounting for the baseline density. This leads to the power estimated from uniform weighted images preserving less of the original diffuse structures.}.}

We subsequently implement a preliminary stepwise foreground subtraction to the images. Bright discrete sources are initially identified in the high-resolution uniform-weighted images using the PyBDSF source-finding algorithm. We filter the source list by spectral index to avoid any misidentification that could remove 21-cm signal or unusual sources. These point sources are then subtracted from the visibilities. Secondly, the diffuse emission, visible prominently in the natural-weighted images, is modelled using WSCLEAN from the point-source-subtracted visibilities. By subtracting both the cataloged point sources and the modelled diffuse emission, we reduce the foreground contamination in the observed sky signal by roughly three orders of magnitude in power distribution and therefore the potential bias from the following 21-cm signal extraction.

Following the preliminary source subtraction, the simulated sky consists of residual foreground, the 21-cm signal and noise. We further process these residual visibilities by re-imaging in the natural deconvolved scheme and then applying Gaussian Process Regression (GPR) to reconstruct the 21-cm signal. The source-subtracted gridded visibility data is modelled with five GPR specifications: \mwedge{}, \anoise{}, \manalytic{}, \avarl{}, \ubase{}. This is conducted within a Bayesian framework using nested sampling for parameter estimation and model comparison. These models explore different parametrizations of the spectral covariance for foreground and the 21-cm signal. The posterior densities of kernel parameters for all the five GPR models show unimodality with well-defined credible intervals. In terms of global evidence, \anoise{} and \mwedge{} are the most competing models for providing the overall description to the simulated sky signals with respect to their highest $\mathrm{log}\mathcal{Z}_M$. We additionally validate the significance of $\mathbf{K}_{21}$ in each model using the same dataset. While all the models demonstrate the preferance of having the 21-cm component, \mwedge{} achieves the highest $\Delta\mathrm{log}\mathcal{Z}_\mathrm{21}$ at $8.01\times10^3$.

We subsequently quantify the local and overall performance of each model in signal recovery using the residuals (\textit{RMSE}) and $z$-scores. \rev{The $z$-score analysis accounts for both the reconstruction accuracy, quantified by the \textit{RMSE}, and the predictive uncertainty arising from the GPR sampling errors in estimating the 21-cm component.} To analyse local trends, we use the per-mode residual and $z$-score to compare the GPR-reconstructed 21-cm signal against the true injected signal at specific $k$-modes. At $k = 0.32\ \mathrm{h\ cMpc}^{-1}$, \anoise{} achieves the best local reconstruction accuracy with a fractional difference of $-0.14\%$, and a $z$-score of $0.315$. In addition, \mwedge{} also reaches its closest recovery at the same $k$-mode, showing a fractional difference of $0.47\%$. In the global analysis, we average the per-mode quantities over all the sampled $k$-modes to yield the overall \textit{RMSE} and \textit{RMS}. From the cross-metric comparisons shown by Figure~\ref{fig:gpr_models_cross_metrics}, \mwedge{} has the lowest \textit{RMSE} of $8.53\ \mathrm{mK^2}$ while \anoise{} achieves the lowest \textit{RMS} $z$-score of $4.37$. Both models effectively reconstruct the underlying cosmological signal with the minimal bias.

In addition, the cross-metric model comparisons above reveals a blind spot in Bayesian model comparisons that only rely on global evidence. Such evidence-based comparisons can favour aggregate models that nonetheless bias individual components. To validate this case, we perform Bayesian null tests to these models by using a dataset that contains no 21-cm signal. Figure~\ref{fig:gpr_models_null_logZ} shows that the validation data correctly prefer the no-21 versions of \mwedge{} and \anoise{} as suggested by their positive $\Delta\mathrm{log}\mathcal{Z}_\mathrm{null}$. The tests therefore suggests these models are robust against spuriously detecting a signal. On the other hand, the results for \ubase{}, \manalytic{} and \avarl{} raise significant concerns. The three models all yield negative $\Delta\mathrm{log}\mathcal{Z}_\mathrm{null}$. This implies that their full models (with the 21-cm component) are preferred even on data containing no 21-cm signal. These behaviours strongly suggest that these models are incorrectly fitting residual noise or foreground structures into their 21-cm component, indicating a high risk of biasing the 21-cm signal estimate when applied to the actual observed data.

% In the reconstruction-based test, \mwedge{}, \anoise{} and \manalytic{} place $1\sigma$ upper limits significantly weaker than the reference 21-cm power and mostly beneath the noise floor.

% The reconstruction test also sees \mwedge{} being the model with the lowest \textit{RMSE} and $z$-score in recovering the signal of primary interest (i.e. the 21-cm signal). \anoise{} and \mwedge{} also achieves the closest local recoveries in fractional difference of $-0.14\%$ and $0.47\%$ between the recovered and injected 21-cm power at $k = 0.32\ \mathrm{h\ cMpc}^{-1}$. 

% To assess detection confidence, we compared each model to a counterpart model with no 21-cm signal kernel component. In our analysis, the GPR models, \mwedge{} and \anoise{} have substantially higher evidence than their counterpart models without 21-cm components. 
% This indicates stronger statistical support for including the 21-cm signal in explaining the observed sky using these two models. 

\section*{Acknowledgements}
\rev{The authors are grateful to the referee for their thoughtful suggestions that improved the manuscript.} YL thanks Dominic Anstey for valuable discussions on Bayesian statistics and Oscar O'Hara and Nicholas Kern for helpful comments on this manuscript. PHS acknowledges the support from
the Science and Technologies Facilities Council. EdLA acknowledges the support of UKRI STFC to this work via an Ernest Rutherford Fellowship.
This work was performed using resources provided by the Cambridge Service for Data Driven Discovery (CSD3) operated by the University of Cambridge Research Computing Service (www.csd3.cam.ac.uk), provided by Dell EMC and Intel using Tier-2 funding from the Engineering and Physical Sciences Research Council (capital grant EP/T022159/1), and DiRAC funding from the Science and Technology Facilities Council (www.dirac.ac.uk).

%%%%%%%%%%%%%%%%%%%%%%%%%%%%%%%%%%%%%%%%%%%%%%%%%%
\section*{Data Availability}
The data underlying this article will be shared upon reasonable request to the corresponding authors.



\bibliographystyle{mnras}
\bibliography{main} % if your bibtex file is called example.bib

%\begin{thebibliography}{99}
%\end{thebibliography}

%%%%%%%%%%%%%%%%%%%%%%%%%%%%%%%%%%%%%%%%%%%%%%%%%%

%%%%%%%%%%%%%%%%% APPENDICES %%%%%%%%%%%%%%%%%%%%%
\clearpage
\appendix

\section{GPR model \rev{parametrization} and prior} \label{sec:gpr_model_parametrization_prior}

In this section, we present the \rev{parametrization} for the five GPR models we analyse in the Bayesian framework discussed above: \mwedge{}, \anoise{}, \avarl{}, \manalytic{} and \ubase{}. Each model is decomposed into four kernel components: intrinsic foregrounds ($\mathbf{K}_\mathrm{int}$), mode-mixed foregrounds ($\mathbf{K}_\mathrm{mix}$), the 21-cm signal ($\mathbf{K}_\mathrm{21}$) and noise ($\mathbf{K}_\mathrm{n}$).
In addition, a prior range is also specified for each kernel parameter used in the nested sampling.

\begin{table}
\centering
\caption{\rev{parametrization} and prior density of kernel hyperparameters in Model \mwedge{}. ``Mat\'ern-5/2'' denotes a Mat\'ern kernel with $\eta = 5/2$, i.e. $\mathbf{K}_\mathrm{Mat\acute{e}rn}(\eta =5/2)$, ``RBF'' denotes a radial basis function, i.e. $\mathbf{K}_\mathrm{Mat\acute{e}rn}(\eta =\infty)$, ``VAE'' denotes the VAE kernel trained in Section~\ref{sec:kernel training} and ``White-Hetero'' denotes a white heteroscedastic kernel described in Section~\ref{sec:candidate_gpr_models}. The noise variance $\sigma^2_\mathrm{n}$ is calibrated against the estimated noise variance and therefore is not assigned a prior range. For any signal component, the variance  $\sigma^2$ is in the same unit as the gridded visibility data in Jansky (Jy) and the length-scale $l$ is in unit of MHz. The wedge angle $\theta_\mathrm{mix}$ is in radians, while other parameters remain dimensionless.} \label{tab:wedge_prior}
\renewcommand{\arraystretch}{1.25}
\begin{tabular}{@{}cccc@{}}
\toprule
\textbf{Component} & \textbf{Kernel} & \textbf{Parameter} & \textbf{Prior} \\
\midrule
\multirow{2}{*}{$\mathbf{K}_\mathrm{int}$}
  & \multirow{2}{*}{Mat\'ern-5/2}
  & $\sigma^2_\mathrm{int}$ 
  & $\mathrm{LogUniform}(-0.6,1.6)$ \\
  & & $l_\mathrm{int}$
  & $\mathrm{Uniform}(10,100)$ \\
\cmidrule{1-4} % (l) (r) (lr)
\multirow{3}{*}{$\mathbf{K}_\mathrm{mix}$}
  & \multirow{3}{*}{RBF}
  & $\sigma^2_\mathrm{mix}$ 
  & $\mathrm{LogUniform}(-4,-0.2)$ \\
  & & $\theta_\mathrm{mix}$ 
  & $\mathrm{Uniform}(0,0.2)$ \\
  & & $C_\mathrm{mix}$ 
  & $\mathrm{Uniform}(0,0.1)$ \\
\cmidrule{1-4}
\multirow{3}{*}{$\mathbf{K}_\mathrm{21}$}
  & \multirow{3}{*}{VAE}
  & $\sigma^2_{21}$ 
  & $\mathrm{LogUniform}(-8,2)$ \\
  & & $x_{1}$ 
  & $\mathrm{Uniform}(-6,6)$ \\
  & & $x_{2}$ 
  & $\mathrm{Uniform}(-6,6)$ \\
\cmidrule{1-4}
\multirow{1}{*}{$\mathbf{K}_\mathrm{n}$}
  & \multirow{1}{*}{White-Hetero}
  & $\sigma^2_\mathrm{n}$ & $-$ \\
\bottomrule
\end{tabular}
\end{table}

\begin{table}
\centering
\caption{\rev{parametrization} and prior density of kernel hyperparameters in Model \anoise{}. The kernels for intrinsic and mode-mixed foregrounds and the 21-cm signal are consistent with those described in Table~\ref{tab:wedge_prior}. In the noise component $\mathbf{K}_\mathrm{n}$, a noise-scaling parameter $\alpha_\mathrm{n}$ is introduced to model variance mismatch in the white heteroscedastic kernel.} \label{tab:alpha_noise_prior}
\renewcommand{\arraystretch}{1.25}
\begin{tabular}{@{}cccc@{}}
\toprule
\textbf{Component} & \textbf{Kernel} & \textbf{Parameter} & \textbf{Prior} \\
\midrule
\multirow{2}{*}{$\mathbf{K}_\mathrm{int}$}
  & \multirow{2}{*}{Mat\'ern-5/2}
  & $\sigma^2_\mathrm{int}$ 
  & $\mathrm{LogUniform}(-0.6,1.6)$ \\
  & & $l_\mathrm{int}$ 
  & $\mathrm{Uniform}(10,100)$ \\
\cmidrule{1-4} % (l) (r) (lr)
\multirow{3}{*}{$\mathbf{K}_\mathrm{mix}$}
  & \multirow{3}{*}{RBF}
  & $\sigma^2_\mathrm{mix}$ 
  & $\mathrm{LogUniform}(-4,-0.2)$ \\
  & & $\theta_\mathrm{mix}$ 
  & $\mathrm{Uniform}(0,0.2)$ \\
  & & $C_\mathrm{mix}$ 
  & $\mathrm{Uniform}(0,0.1)$ \\
\cmidrule{1-4}
\multirow{3}{*}{$\mathbf{K}_\mathrm{21}$}
  & \multirow{3}{*}{VAE}
  & $\sigma^2_{21}$ 
  & $\mathrm{LogUniform}(-8,2)$ \\
  & & $x_{1}$ 
  & $\mathrm{Uniform}(-6,6)$ \\
  & & $x_{2}$ 
  & $\mathrm{Uniform}(-6,6)$ \\
\cmidrule{1-4}
\multirow{2}{*}{$\mathbf{K}_\mathrm{n}$}
  & \multirow{2}{*}{White-Hetero}
  & $\sigma^2_\mathrm{n}$ 
  & $-$ \\
  & & $\alpha_\mathrm{n}$ 
  & $\mathrm{Uniform}(0.8,1.2)$ \\
\bottomrule
\end{tabular}
\end{table}

\begin{table}
\centering
\caption{\rev{parametrization} and prior density of kernel hyperparameters in Model \avarl{}. The kernels for intrinsic foregrounds and the 21-cm signal are consistent with those described in Table~\ref{tab:wedge_prior}. The mode-mixed component $\mathbf{K}_\mathrm{mix}$ is reparameterized with two power indices $\alpha^\mathrm{var}_\mathrm{mix}$ and $\alpha^{l}_\mathrm{mix}$ that scales $\sigma^2_\mathrm{mix}$ and $l_\mathrm{mix}$ with baseline length.} \label{tab:alpha_var_l_prior}
\renewcommand{\arraystretch}{1.25}
\begin{tabular}{@{}cccc@{}}
\toprule
\textbf{Component} & \textbf{Kernel} & \textbf{Parameter} & \textbf{Prior} \\
\midrule
\multirow{2}{*}{$\mathbf{K}_\mathrm{int}$}
  & \multirow{2}{*}{Mat\'ern-5/2}
  & $\sigma^2_\mathrm{int}$ 
  & $\mathrm{LogUniform}(-0.6,1.6)$ \\
  & & $l_\mathrm{int}$ 
  & $\mathrm{Uniform}(10,100)$ \\
\cmidrule{1-4} % (l) (r) (lr)
\multirow{4}{*}{$\mathbf{K}_\mathrm{mix}$}
  & \multirow{4}{*}{RBF}
  & $\sigma^2_\mathrm{mix}$ 
  & $\mathrm{LogUniform}(-4,-0.2)$ \\
  & & $\alpha^\mathrm{var}_\mathrm{mix}$ 
  & $\mathrm{Uniform}(-2,0)$ \\
  & & $l_\mathrm{mix}$ 
  & $\mathrm{Uniform}(2,6)$ \\
  & & $\alpha^{l}_\mathrm{mix}$ 
  & $\mathrm{Uniform}(0.5,1.5)$ \\
\cmidrule{1-4}
\multirow{3}{*}{$\mathbf{K}_\mathrm{21}$}
  & \multirow{3}{*}{VAE}
  & $\sigma^2_{21}$ 
  & $\mathrm{LogUniform}(-8,2)$ \\
  & & $x_{1}$ 
  & $\mathrm{Uniform}(-6,6)$ \\
  & & $x_{2}$ 
  & $\mathrm{Uniform}(-6,6)$ \\
\cmidrule{1-4}
\multirow{1}{*}{$\mathbf{K}_\mathrm{n}$}
  & \multirow{1}{*}{White-Hetero}
  & $\sigma^2_\mathrm{n}$ & $-$ \\
\bottomrule
\end{tabular}
\end{table}

\begin{table}
\centering
\caption{parametrization and prior density of kernel hyperparameters in Model \manalytic{}. The kernels for intrinsic and mode-mixed foregrounds are consistent with those described in Table~\ref{tab:wedge_prior}. In this model, the 21-cm component is parametrized with an analytic exponential kernel (EXP), $\mathbf{K}_\mathrm{Mat\acute{e}rn}(\eta = 1/2)$.} \label{tab:analytic_prior}
\renewcommand{\arraystretch}{1.25}
\begin{tabular}{@{}cccc@{}}
\toprule
\textbf{Component} & \textbf{Kernel} & \textbf{Parameter} & \textbf{Prior} \\
\midrule
\multirow{2}{*}{$\textbf{K}_\mathrm{int}$}
  & \multirow{2}{*}{Mat\'ern-5/2}
  & $\sigma^2_\mathrm{int}$ 
  & $\mathrm{LogUniform}(-0.6,1.6)$ \\
  & & $l_\mathrm{int}$ 
  & $\mathrm{Uniform}(10,100)$ \\
\cmidrule{1-4} % (l) (r) (lr)
\multirow{3}{*}{$\textbf{K}_\mathrm{mix}$}
  & \multirow{3}{*}{RBF}
  & $\sigma^2_\mathrm{mix}$ 
  & $\mathrm{LogUniform}(-4,-0.2)$ \\
  & & $\theta_\mathrm{mix}$ 
  & $\mathrm{Uniform}(0,0.2)$ \\
  & & $C_\mathrm{mix}$ 
  & $\mathrm{Uniform}(0,0.1)$ \\
\cmidrule{1-4}
\multirow{2}{*}{$\textbf{K}_\mathrm{21}$}
  & \multirow{2}{*}{EXP}
  & $\sigma^2_{21}$ 
  & $\mathrm{LogUniform}(-8,2)$ \\
  & & $l_\mathrm{21}$ 
  & $\mathrm{Uniform}(0.1,0.25)$ \\
\cmidrule{1-4}
\multirow{1}{*}{$\mathbf{K}_\mathrm{n}$}
  & \multirow{1}{*}{White-Hetero}
  & $\sigma^2_\mathrm{n}$ & $-$ \\
\bottomrule
\end{tabular}
\end{table}

\begin{table}
\centering
\caption{parametrization and prior density of kernel hyperparameters in Model \ubase{}. The kernel for the 21-cm component is consistent with that described in Table~\ref{tab:wedge_prior}. The kernel for each foreground component is reparameterized to a single variance $\sigma^2$ and a single length-scale $l$, independent of baselines.} \label{tab:unibaseline_prior}
\renewcommand{\arraystretch}{1.25}
\begin{tabular}{@{}cccc@{}}
\toprule
\textbf{Component} & \textbf{Kernel} & \textbf{Parameter} & \textbf{Prior} \\
\midrule
\multirow{2}{*}{$\textbf{K}_\mathrm{int}$}
  & \multirow{2}{*}{Mat\'ern-5/2}
  & $\sigma^2_\mathrm{int}$ 
  & $\mathrm{LogUniform}(-0.6,1.6)$ \\
  & & $l_\mathrm{int}$ 
  & $\mathrm{Uniform}(10,100)$ \\
\cmidrule{1-4} % (l) (r) (lr)
\multirow{2}{*}{$\textbf{K}_\mathrm{mix}$}
  & \multirow{2}{*}{RBF}
  & $\sigma^2_\mathrm{mix}$ 
  & $\mathrm{LogUniform}(-4,-0.2)$ \\
  & & $l_\mathrm{mix}$ 
  & $\mathrm{Uniform}(2,6)$ \\
\cmidrule{1-4}
\multirow{3}{*}{$\textbf{K}_\mathrm{21}$}
  & \multirow{3}{*}{VAE}
  & $\sigma^2_{21}$ 
  & $\mathrm{LogUniform}(-8,2)$ \\
  & & $x_{1}$ 
  & $\mathrm{Uniform}(-6,6)$ \\
  & & $x_{2}$ 
  & $\mathrm{Uniform}(-6,6)$ \\
\cmidrule{1-4}
\multirow{1}{*}{$\mathbf{K}_\mathrm{n}$}
  & \multirow{1}{*}{White-Hetero}
  & $\sigma^2_\mathrm{n}$ & $-$ \\
\bottomrule
\end{tabular}
\end{table}

\clearpage
\section{Posterior density distribution}
This appendix presents the posterior distributions of the kernel parameters for the GPR models we use in this work, estimated from nested sampling. \label{sec:gpr_model_parametrization_posterior}

\begin{figure*}
    \centering
    \includegraphics[width=\linewidth]{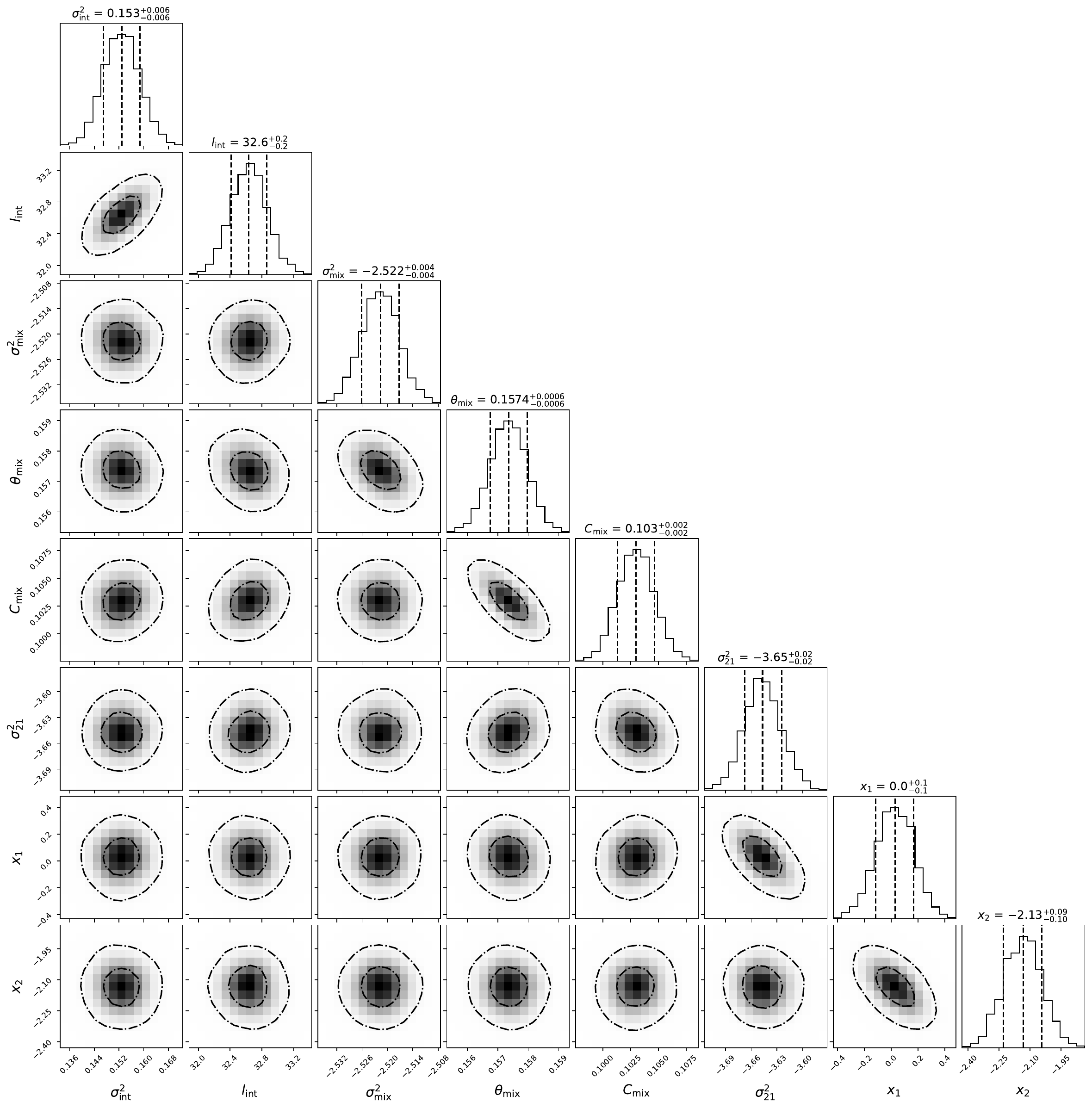}
    \caption{Posterior contours showing the 68\% and 95\% credible regions and the 1D marginals for \mwedge{}. The 16\%, 50\% and 84\% quantiles are marked on the 1D marginals; the median (50\%) is also indicated for each kernel parameter as the nested sampling estimate with the lower (16\%) and upper uncertainty (84\%). These density distributions show unimodality with well-defined boundaries. The variances of the intrinsic and mode-mixed foregrounds, $\sigma^2_\mathrm{int}$ and $\sigma^2_\mathrm{mix}$, and the 21-cm variance $\sigma^2_{21}$ exhibit weak correlations, indicating that these components are cleanly separated by the model from the observed data. The wedge parameters, $\theta_\mathrm{mix}$ and $C_\mathrm{mix}$, are slightly correlated, as expected for the angle-buffer trade-off.} \label{fig:wedge_posterior}
\end{figure*}

\begin{figure*}
    \centering
    \includegraphics[width=\linewidth]{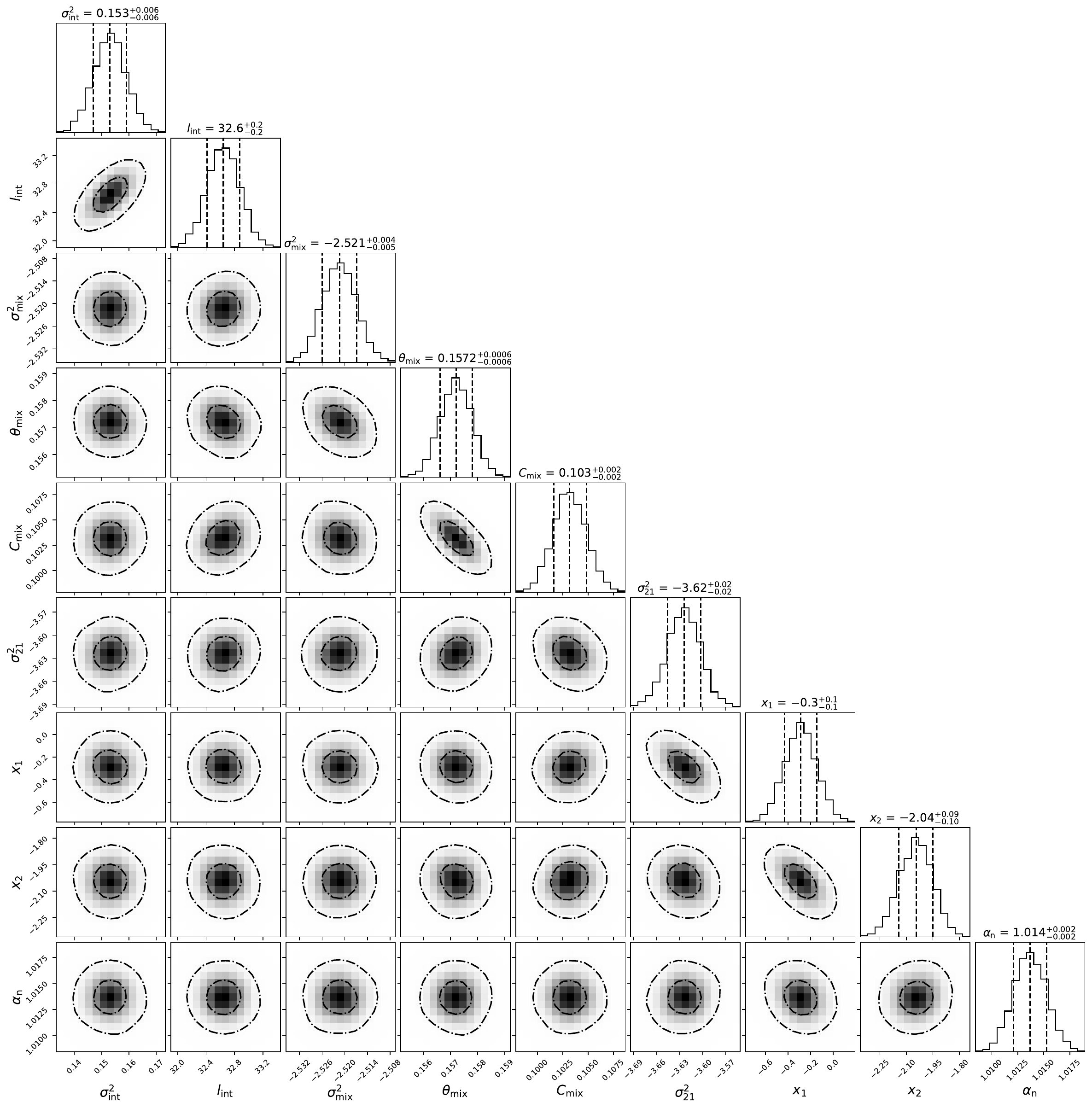}
    \caption{Posterior contours showing the 68\% and 95\% credible regions and the 1D marginals for \anoise{}; the 16\%, 50\%, and 84\% quantiles are also marked on the 1D marginals. Similar to Model \mwedge{}, the posteriors are unimodality and well-constrained. The foreground and 21-cm amplitudes, described by  $\sigma^2_\mathrm{int}$, $\sigma^2_\mathrm{mix}$ and $\sigma^2_{21}$, remain weakly correlated. The lack of pronounced degeneracy implies a clear separation of the well-defined components. The noise-scaling parameter, $\alpha_\mathrm{n}$, is also tightly constrained and only weakly coupled to the other components, suggesting that the noise is identified independently rather than absorbed by the foreground or 21-cm terms.} \label{fig:alpha_noise_posterior}
\end{figure*}

\begin{figure*}
    \centering
    \includegraphics[width=\linewidth]{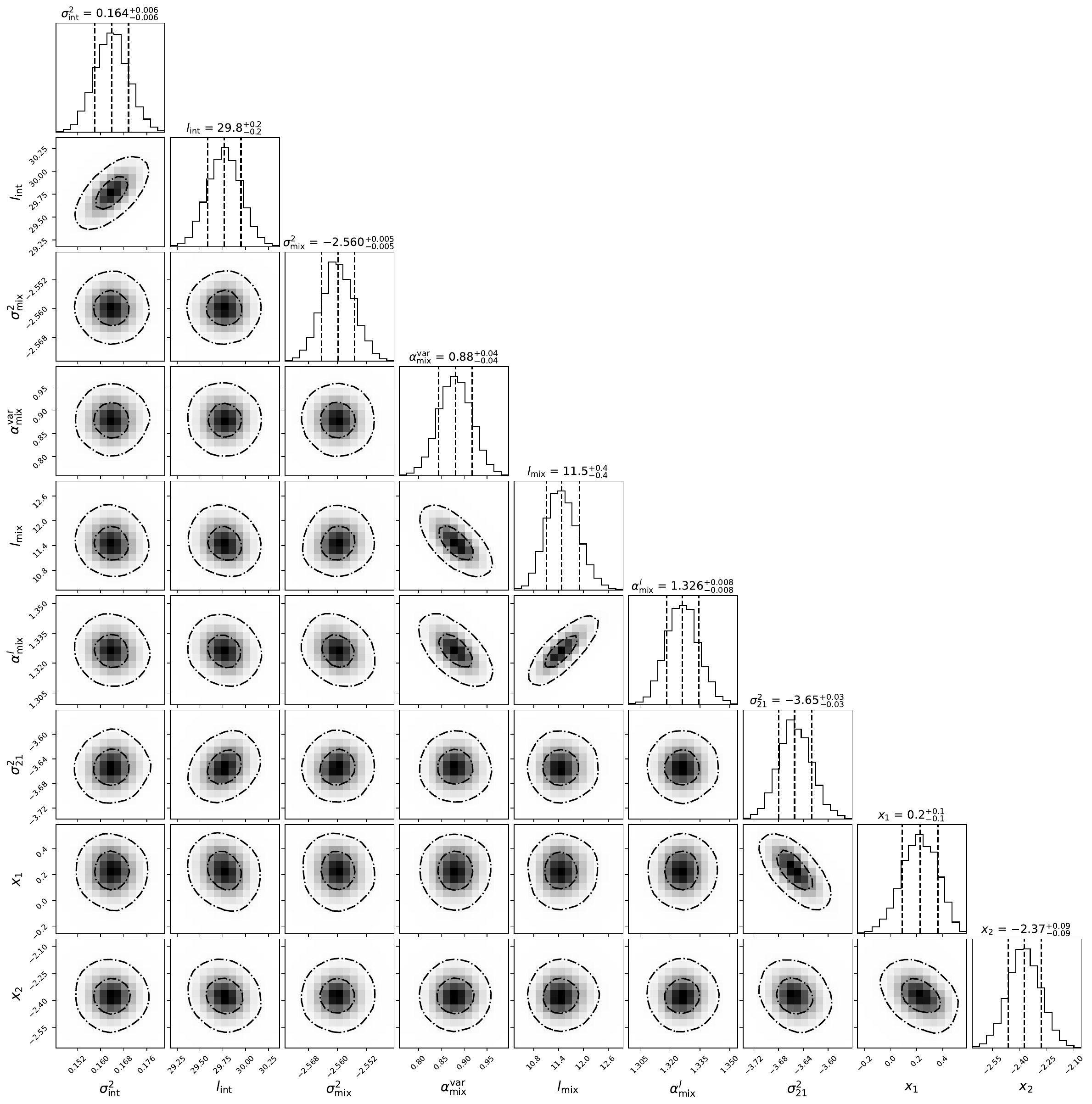}
    \caption{Posterior contours showing the 68\% and 95\% credible regions and the 1D marginals for \avarl{}; the 16\%, 50\%, and 84\% quantiles are also marked on the 1D marginals. Besides showing unimodality, the signal variance, $\sigma^2_\mathrm{int}$, $\sigma^2_\mathrm{mix}$ and $\sigma^2_{21}$ are distinctly separated from each other. While a weak correlation is seen between $\sigma^2_\mathrm{mix}$ and the mode-mix length-scale $l_\mathrm{mix}$, the variance-scaling index $\alpha^\mathrm{var}_\mathrm{mix}$ is found more correlated with both $\alpha^\mathrm{var}_\mathrm{mix}$ and $\alpha^{l}_\mathrm{mix}$. Despite being well bounded, the posterior shows a more noticeable degeneracy between the variation of baseline-dependent amplitude and coherence length scaling when describing the mode-mixed foreground across $u$-space with this model.} \label{fig:alpha_var_l_posterior}
\end{figure*}

\begin{figure*}
    \centering
    \includegraphics[width=\linewidth]{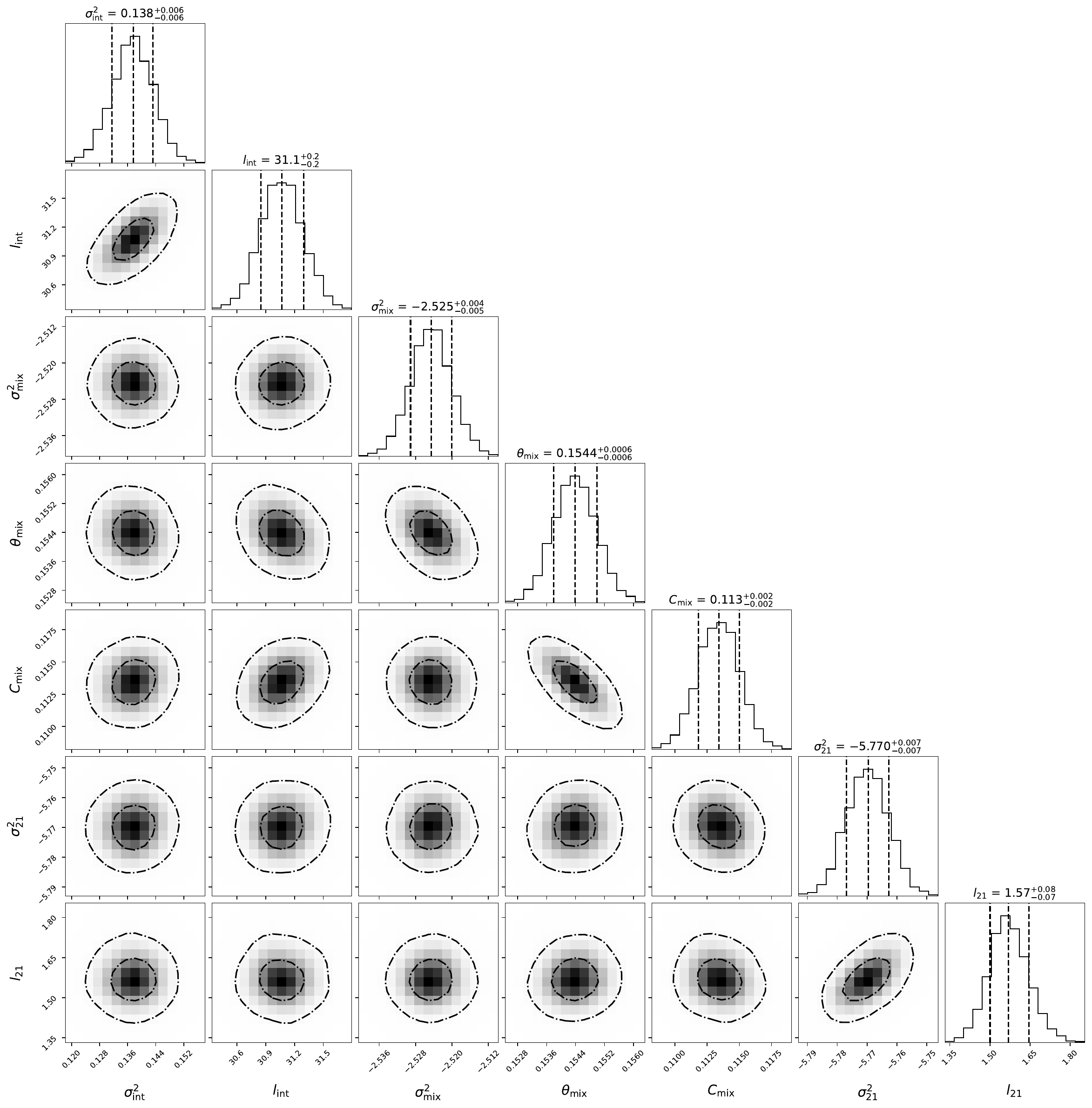}
    \caption{Posterior contours showing the 68\% and 95\% credible regions and the 1D marginals for \manalytic{}; the 16\%, 50\%, and 84\% quantiles are also marked on the 1D marginals. With the exponential Mat\'ern kernel, the model predicts a much weaker 21-cm signal than the VAE models, at $\sigma^2_{21} = -5.767^{+0.007}_{-0.007}$. Meanwhile, the 21-cm kernel parameters, $\sigma^2_{21}$ and $l_{21}$, exhibit a slight positive correlation. Despite being in lower amplitude, $\sigma^2_{21}$ remain well constrained and weakly coupled to the foreground components, $\sigma^2_\mathrm{int}$ and $\sigma^2_\mathrm{mix}$. However, due to the same modelling of mode-mixed foregrounds as \mwedge{} and \anoise{}, the choice of this analytic kernel does not significantly bias the values of the mode-mixing variance $\sigma^2_\mathrm{mix}$, wedge angle $\theta_\mathrm{mix}$ and buffer constant $C_\mathrm{mix}$.} \label{fig:analytic_posterior}
\end{figure*}

\begin{figure*}
    \centering
    \includegraphics[width=\linewidth]{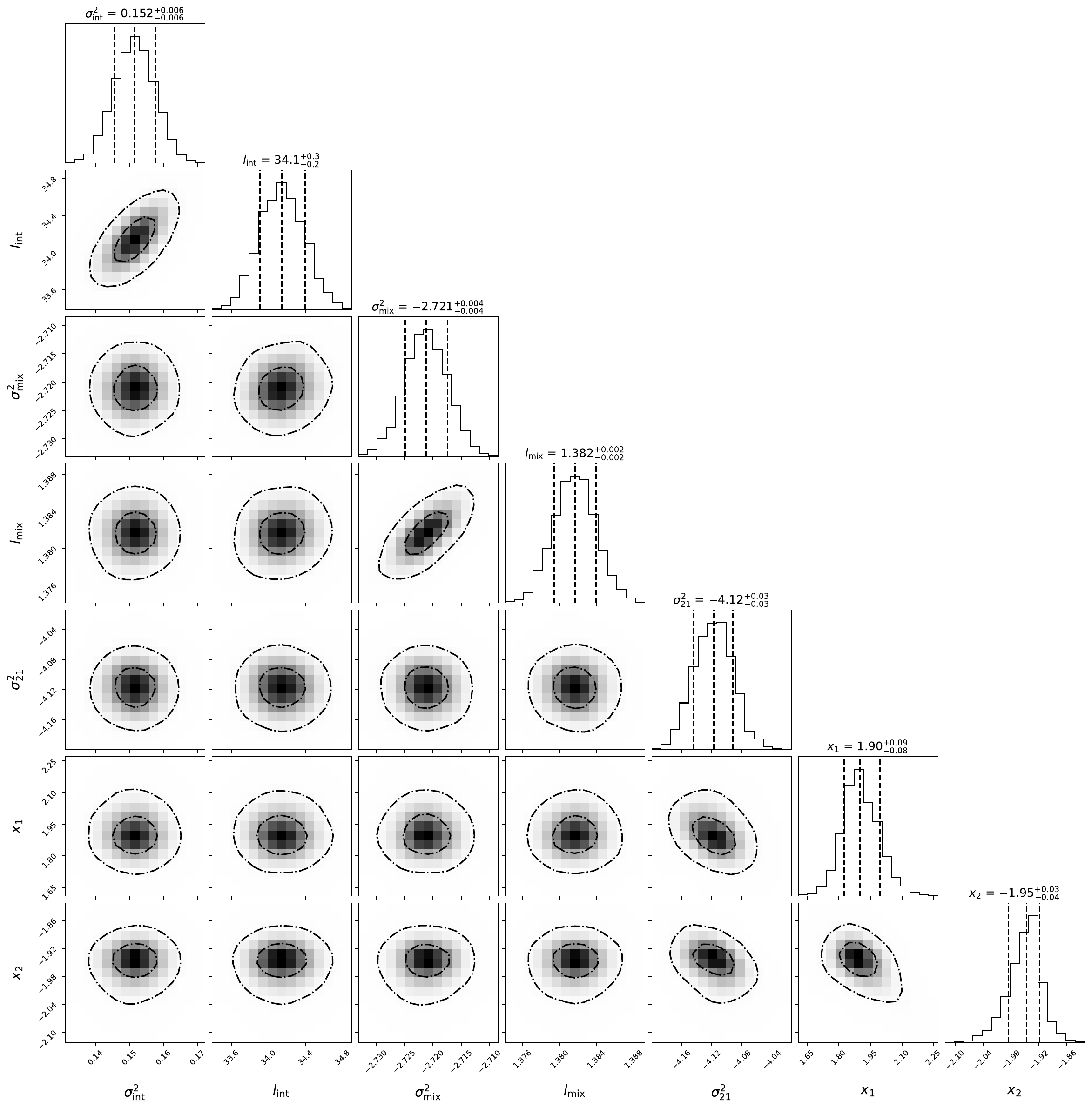}
    \caption{Posterior contours showing the 68\% and 95\% credible regions and the 1D marginals for \ubase{}; the 16\%, 50\%, and 84\% quantiles are also marked on the 1D marginals. The posteriors are unimodal, consistent with the other GPR models. A noticeable degeneracy is seen while imposing the baseline-independent parameters for the mode-mixed foreground. The amplitude $\sigma^2_\mathrm{mix}$ shows a negative correlation with the coherence length $l_\mathrm{mix}$ under this baseline-independent parametrization. Meanwhile, the posterior for the 21-cm variance $\sigma^2_{21}$ shifts to a lower regime compared to the multi-baseline VAE models; the reduction therefore leads to a slightly weaker reconstructed 21-cm power.} \label{fig:unibaseline_posterior}
\end{figure*}

%%%%%%%%%%%%%%%%%%%%%%%%%%%%%%%%%%%%%%%%%%%%%%%%%%

% Don't change these lines
\bsp	% typesetting comment
\label{lastpage}
\end{document}